# Intracellular luminescence thermometry: A story of disagreement, trust, and hope.

Araceli de Aquino Samper,[a] Liyan Ming,[a] Daniel Jaque,[b,c] Riccardo Marin[a,c*]

[a] *Intelligent Optical Nanomaterials (IONs) Lab, Departamento de Química Orgánica, Facultad de Ciencias, Universidad Autónoma de Madrid, 28049 Madrid, Spain*

[b] *Nomaterials for BioImaging Group (nanoBIG), Departamento de Física de Materiales, Facultad de Ciencias, Universidad Autónoma de Madrid, 28049 Madrid, Spain*

[c] *Institute for Advanced Research in Chemistry (IAdChem), 28049 Madrid, Spain*

## Table of Contents

## Abstract

Intracellular luminescence thermometry has long promised to reveal how heat is generated, dissipated, and regulated inside living cells. Yet, despite substantial progress, the field remains shaped by disagreement over the magnitude and physical plausibility of reported intracellular temperature gradients. In this manuscript, we discuss luminescence thermometry as a powerful approach for probing temperature at subcellular length scales, while emphasizing the experimental care required to make such measurements meaningful. After outlining the field's development, we outline the relevant heat-transfer concepts, before introducing luminescence thermometry and the performance metrics used to describe precision and accuracy. We then examine how thermometer design, intracellular localization, calibration, microscopy configuration, and data treatment influence the final thermal readout. Particular attention is given to two recurrent sources of error: bias, arising from measurement conditions and optical distortions, and cross-sensitivity, arising when the probe responds to parameters other than temperature, such as pH, viscosity, ionic strength, or biomolecular interactions. Finally, we outline practical directions for improving reproducibility, including multi-feature readouts, machine-learning-assisted analysis, and FAIR data practices, while suggesting future research directions.[1,2,3]

## 1. Introduction

Temperature is a key parameter in cellular physiology. On the one hand, metabolic activity generates heat: a process whose magnitude is cell-dependent, with adipocytes and tumour cells showing marked thermogenic capabilities.[4–6] On the other hand, the physiology of cells is modulated by external thermal variations, which have impact, among other parameters, on enzymatic activity and membrane fluidity.[7,8] The capability of monitoring intracellular temperature would hence deepen our understanding of life's fundamental building blocks, as well as help understanding disruptive diseases and driving the development of therapeutic approaches.[9,10,11] Among the approaches proposed for intracellular temperature monitoring, luminescence thermometry is considered particularly attractive.

Luminescence thermometry relies on optical probes such as molecular dyes, polymeric nanoparticles, semiconductor nanocrystals, lanthanide-doped nanoparticles, nanodiamonds, and genetically encoded proteins, whose luminescence is modulated by temperature changes.[1,12–14] Optical signals—such as emission intensity, intensity ratio, lifetime, and spectral shift—can be calibrated as thermometric readout features and subsequently monitored to obtain intracellular temperature maps with submicrometric spatial resolution.

Compared with other approaches, luminescence thermometry overcomes several limitations. Traditional contact thermometers become ineffective when probing systems below 10 μm and are invasive.[3,15] Despite some examples of applications for cell studies,[16] infrared (IR) thermometry—aside from being limited to measuring surface temperatures—is still an emergent technology and is intrinsically characterized by a reduced, diffraction-limited spatial resolution, allowing for a coarse-grained view of cells. Raman-based thermometry, albeit capable of working label-free, often suffers from weak signals and spectral crosstalk with fluorescent species.[17,18] Average cell temperature can be retrieved via micro-thermocouples and microelectrodes, yet at the cost of losing mapping capabilities and high invasiveness.[19] Another rather specialized label-free approach entails monitoring temperature-induced refractive index changes in the medium,[20,21] yet with limited applicability at the cellular level—given the marked variability of this parameter due to the dynamic nature of the cytoplasm—and a minimum detectable temperature change in the order of 1 ºC. In contrast, luminescence thermometry offers a minimally-invasive approach with sub-second response times and sub-micrometric spatial resolution, often capable of resolving temperature changes down to 0.1 ºC.[14,22]

The development of intracellular luminescence thermometry (**Figure 1**) has been driven by interdisciplinary collaboration between physicists and engineers, who design advanced microscopy instrumentation; chemists, who synthesize and optimize fluorescent probes; and biologists, who investigate the functions and dynamics of living systems. Since the invention of confocal microscopy by Marvin Minsky in 1955, subcellular imaging techniques have undergone continuous evolution. Confocal microscopy laid the foundation for modern subcellular imaging by using a focused laser beam to raster scan samples and a pinhole to reject unfocused light. Initially developed for micro-Raman studies,[23] this configuration enables optical sectioning and high-resolution imaging of biological specimens, and it has served as the basis for numerous laser scanning microscopy techniques tailored to specific scientific applications.[24]

Despite these remarkable technological advances, the scientific community remains fragmented: Each laboratory typically uses its own thermometers, calibration procedures, microscope settings, and analysis scripts. Although the resolution of modern probes can reach 0.1 °C under ideal conditions, it is experimentally difficult to achieve, and our experience tells us that its reproducibility is often questionable. In addition, the discrepancies between the measured values can vary by several degrees between studies.[3,25,26]

The appearance of the first reports on intracellular temperature mapping through fluorescence thermometry set a vigorous debate in motion (**Figure 1**). Baffou and co-workers have always been adamant about the impossibility that thermal gradients can be sustained at the intracellular level (unless artificially induced via, e.g., localized photothermal processes), strong of theoretical arguments and the results reported in their seminar work using green fluorescence protein (GFP) as a polarization-based luminescent thermometer.[27] Moving from these considerations, they have questioned the physical plausibility of the large intracellular temperature gradients reported in experimental studies: A "*$10^5$ gap*", as dubbed in the authors.[27] They state that, in the context of normal metabolic activity and realistic sugar concentration levels, such gradients are not physically and biologically credible. They also suggested that the apparent temperature increases observed via luminescence thermometry could originate from optical artifacts rather than genuine heating effects. For example, they pointed out that fluorescence variations caused by dye aggregation or photobleaching could be misinterpreted as temperature changes, while non-uniform excitation intensity across the imaging field and the lack of appropriate in-cell calibration procedures could further contribute to inaccurate measurements.

In response to Baffou, Uchiyama and colleagues[10,28]—experts in polymer and lifetime-based thermometry—defended the reliability of intracellular temperature sensing. They argued that Baffou's theoretical objections had not been experimentally validated in biological systems and pointed to extensive experimental evidence supporting the validity of intracellular thermometry. This included consistent results obtained across different cell lines using the same thermometer, as well as within the same cell line using thermometers based on different sensing mechanisms. They concluded by emphasizing the need for further experimental and theoretical studies to reconcile the observed data with existing thermodynamic models, while also acknowledging that the interpretation of fluorescence changes remains complex and non-trivial.

Further support to the observations of the group of Uchiyama came in 2018, when Chrétien et al.[29] reported that mitochondria could operate close to 50 °C while the surrounding cytosol remained near 37 °C. These results, obtained using the fluorescent probe Mito Thermo Yellow (MTY), suggested that the cell environment might sustain temperature gradients of ≈10 °C over submicron distances. Yet, critics pointed out that MTY is also sensitive to local changes in viscosity, and hence the reported value of temperature can be overestimated.

Subsequent works introduced improved calibration strategies and lifetime-based thermometric readouts designed to minimize interference in the readout introduced by the highly dynamic and heterogeneous cell environment as well as by the optical setup. To that end, in 2021, Okabe and Uchiyama[30] emphasized that reliable intracellular thermometry depends critically on the calibration process, which must reproduce as closely as possible the physicochemical conditions of the intracellular environment. The need to compare the results of calibrations run in different conditions to guarantee accurate thermal readout started therefore to become evident in the field. This argument is also corroborated by the work of Rodríguez-Sevilla et al., who demonstrated in 2023 how intracellular thermal readout using GFP as luminescent thermometer can lead to erroneous readouts owing to the protein interaction with other chemical species in the cell's compartments and the different environment in live and fixed cells.

Over the past 4-5 years, a less conventional perspective has emerged in luminescence thermometry broadly speaking, that employs advanced data analysis approaches—including multivariate analysis and machine learning (ML) algorithms—to analyse the calibration dataset for enhanced temperature readout precision and accuracy. To that end, Ximendes, Marin, Carlos and Jaque[31] approached calibration from a data-science perspective. They demonstrated that the spectral datasets collected during the calibration process contain a high degree of redundant information that can be statistically optimized in an automated way. By applying dimensionality reduction techniques such as Principal Component Analysis (PCA) and t-distributed Stochastic

Neighbor Embedding (t-SNE)—broadly falling under the ML algorithms category—the authors improved the thermal resolution ($\delta T$). More recently, ML-driven approaches have also been extended to solve issues related with readout interference of off-target parameter changes (i.e., cross-sensitivity), hence hinting at a powerful tool to render luminescence thermometry reliable also in highly complex biological environments.[32–34] Despite the fact that these approaches have yet to be demonstrated in intracellular thermometry, we believe they hold an immense potential in cell studies.

Together, these studies highlight a conceptual shift in luminescence thermometry: calibration is no longer limited to the experimental adjustment of probe response but extends to the computational optimization of the dataset itself. This evolution further reinforces the need for methodological standards that integrate both experimental and analytical best practices to ensure accuracy, reproducibility, and cross-comparability in nanoscale temperature measurements, while also providing high-quality datasets for reliably training ML algorithms.

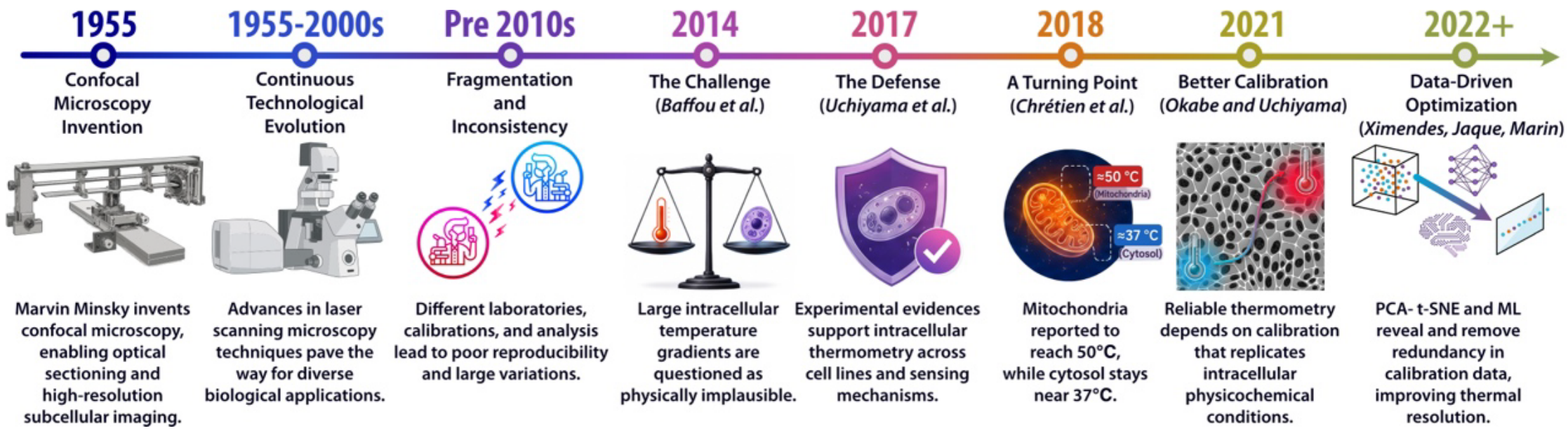

**Figure 1.** Timeline of the intracellular thermometry including the "*$10^5$ gap*" debate.

The main contentious points preventing consensus to be achieved inside the field can be summarized as: ***i)*** **Physical constraints**: Some researchers claim that the magnitudes of heat generation in biologically-realistic metabolic activity and Kapitza resistance in biological membranes are incompatible with measurable thermal gradients. ***ii)*** **Interference**: a variety of factors can distort the measured signal, including photobleaching, local excitation heterogeneity, scattering, self-absorption, laser-induced heating, variability in the local concentration of sensors, or lack of consistency in the used optical setup (a class of phenomena collectively referred to as bias). Biological and chemical factors such as cell movement, changes in refractive index, as well as changes in pH, ionic strength, viscosity, or protein binding can modify the fluorescence emission independently of temperature (a class of phenomena collectively referred to as cross-sensitivity). ***iii)*** **Data-processing:** the lack of standardized analysis pipelines, metadata reporting, and FAIR-data sharing hinders reproducibility and cross-comparison. ***iv)*** **Lack of commercial thermometers and validation experiments**: due to the issues illustrated above, currently there is no credible candidate material to become part of a commercial temperature measurement kit (unlike for, e.g., $Ca^{2+}$ and reactive oxygen species, ROS). Moreover, no standard operating procedure (SOP) is available for guaranteeing reliability of the thermometric approach, which represents an unsatisfied precondition for the development of such a commercial kit.

Current evidence suggest that ***the field now requires a coherent and consensual experimental framework*** that integrates thermal physics, optical metrology, and biological validation. This manuscript identifies the design principles of a luminescent sensor for intracellular luminescence thermometry, also describing how to calibrate it and use it to carry out precise and accurate measurements in cells. Given the lack of consensus in the literature, we do not propose this as the "ultimate" guide to intracellular luminescence thermometry. Rather, this article represents both a primer and an ongoing effort to standardize the field of intracellular luminescence thermometry and set the foundations to address unanswered biological questions in a (more) reliable manner. To this end, the article is organized into three main sections: ***i)*** the theoretical framework and core concepts underlying reliable thermometers, ***ii)*** experimental and analytical considerations, including instrumentation, calibration, and measurement procedures, and ***iii)*** post-analysis aspects, encompassing artifact recognition, data analysis, FAIR practices, validation, and future perspectives.

# 2. Core concepts

## 2.1. Readout features

Luminescence signals that can be used as readout features (**Figure 2** and **Table 1**) include[14] *i)* absolute or relative variations in the integrated emission intensity of one or multiple bands (a-c), *ii)* spectral modifications such as shifts in emission peak position or bandwidth (d, e), and *iii)* changes in the time domain (f).

The nature of the biological system or process under study, together with the used readout method, dictates the required working temperature range, spatial resolution, and overall measurement conditions. Selecting or designing an appropriate luminescent probe relies heavily on weighing the advantages and limitations of each readout approach.[2,3,25,35]

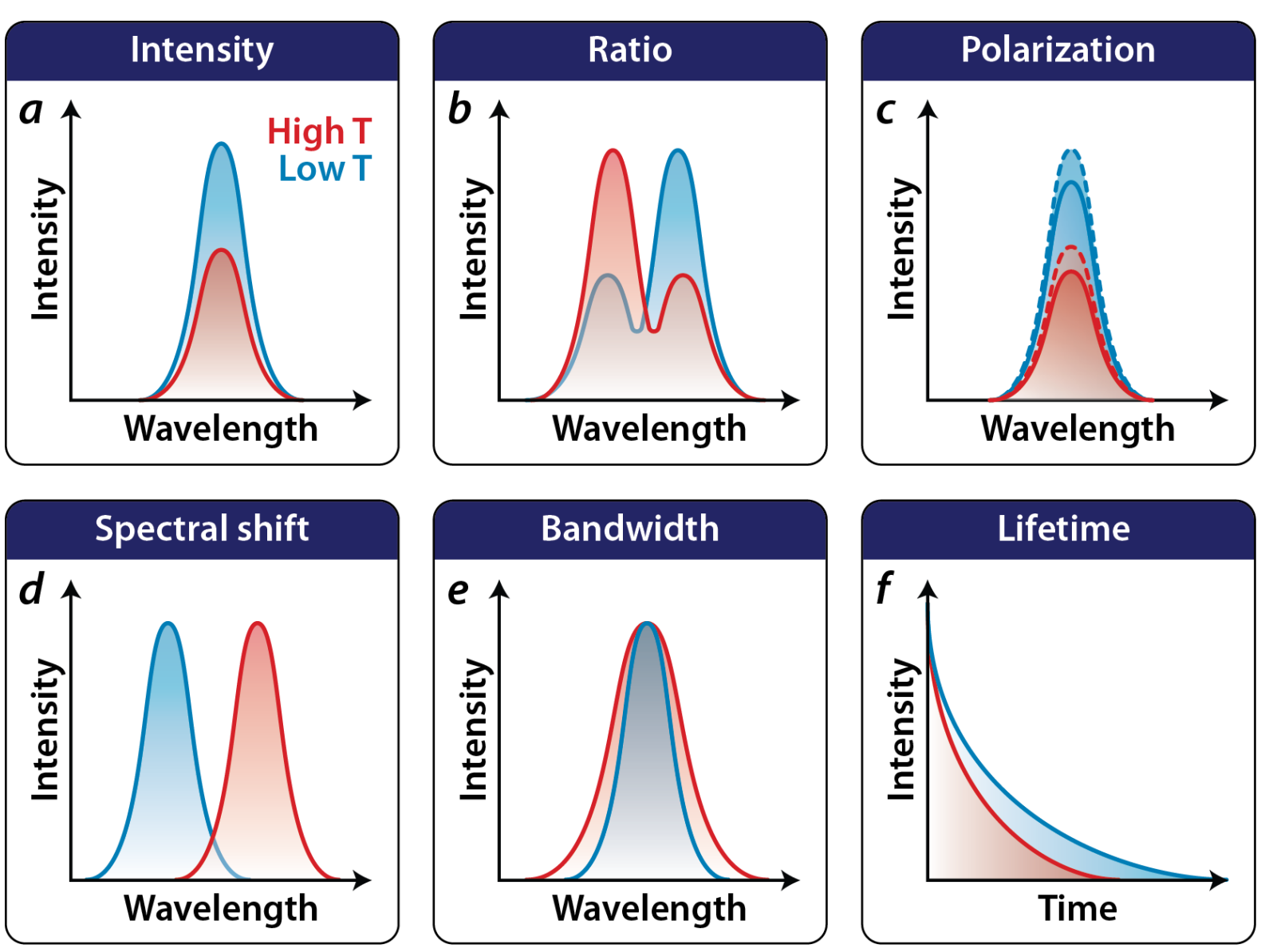

**Figure 2.** Schematic comparison of the possible readout features when sensing temperature (red: high T, blue: low T). Image created based on a published work.[3]

**Table 1.** Comparison of the vulnerabilities and technological requirements of the described luminescent readouts.

| READOUT FEATURE \ INTERFERENCE SOURCE | Change in excitation intensity | Tirrue-photon interaction | Change in probe concentration | Self-absorption | Environmental changes (pH, viscosity, ions...) |
|---|---|---|---|---|---|
| Intensity | xxx | xxx | xxx | xxx | xxx |
| Ratio | x[a] | xx | x | xx | xxx |
| Polarization | x | xx | x | x | xxx |
| Spectral shift/ Bandwidth | x | xx | x | xx | xxx |
| Lifetime | x[a] | x[b] | x | x | xxx |

*level of severity:* **x** *Low*, **xx** *Medium*, **xxx** *High*

[a] only if using bands with different photon order

[b] only for ns lifetime, scales with optical path

While absolute intensity measurements are technically straightforward and fast, they are highly sensitive to experimental and environmental fluctuations. To overcome these limitations, alternative strategies, such as ratiometric detection (luminescence intensity ratio - LIR), polarization-based readouts, spectral/bandwidth tracking, and lifetime-based thermometry are frequently employed. Each method offers a different combination of accuracy, structural immunity to readout interferences, and instrumental complexity and cost.[11,36,37] The specific vulnerabilities and technological requirements of these primary luminescent readout methods are cross-compared in the table below.

### 2.1.1. Multimodal/multi-feature combinations[31,38,32,39]

Multimodal/multi-feature thermometry represents an alternative strategy to improve the reliability of temperature measurements. In this case, rather than relying on a single readout feature, several independent readout features can be acquired and analysed either individually or in combination. For instance, Suchocki et al.[40] demonstrated that combining different readout features, including spectral ratios and lifetime readouts, within a single probe enables internal consistency checks between independent temperature estimates.

One of the main advantages of this approach is that, given a luminescent probe, different readout features can be affected by different sources of errors. Therefore, agreement among multiple readouts provides and internal consistency check. This concept is particularly relevant for intracellular thermometry, where changes in pH, viscosity, ionic strength, or biomolecular interactions may affect individual readouts differently.

### 2.1.2. Machine learning derived readouts

Beyond classical readout features and with the advent of ML algorithms, we have observed the introduction of readout features that go beyond the classical spectral features described above. Classical spectral features have an intrinsic "local" nature, meaning that they describe a specific element of the emission spectrum (e.g., the position of the band maximum or the relative change in the intensity over two wavelength ranges). However, a more holistic, "global" approach has been proposed lately, which leverages the capability of ML algorithms to identify patterns in high-dimensionality datasets. Dimensionality reduction (DR) and neural network-based algorithms, specifically, have been implemented to identify, from the analysis of the whole emission spectrum, linear and non-linear combinations of intensities to unlock readout features with a more complex spectral structure and enhanced thermometric performance. Even though these readout features carry intrinsic memory of the temperature-dependent physical processes, the algorithms generally do not explicitly use this knowledge during the training. Take the case of, DR algorithms such as principal component analysis (PCA) and linear discriminant analysis (LDA), which identify linear combination of spectral features that are effectively projections of experimental spectra and that can be used as readout features (**Figure 3**).[31,32] The values of the coefficients defining the relative weight of the intensity at each wavelength highlight wavelength ranges in which spectral changes as a function of the target parameter (here, temperature) are more relevant—i.e., they describe a high fraction of the dataset's variance. Neural network-based ML algorithms are less revealing in that sense, often relying on mathematical constructs to harness non-linear. In all cases, ML-extracted readout features should be treated with extreme care, given the proclivity to overfitting with this approach.

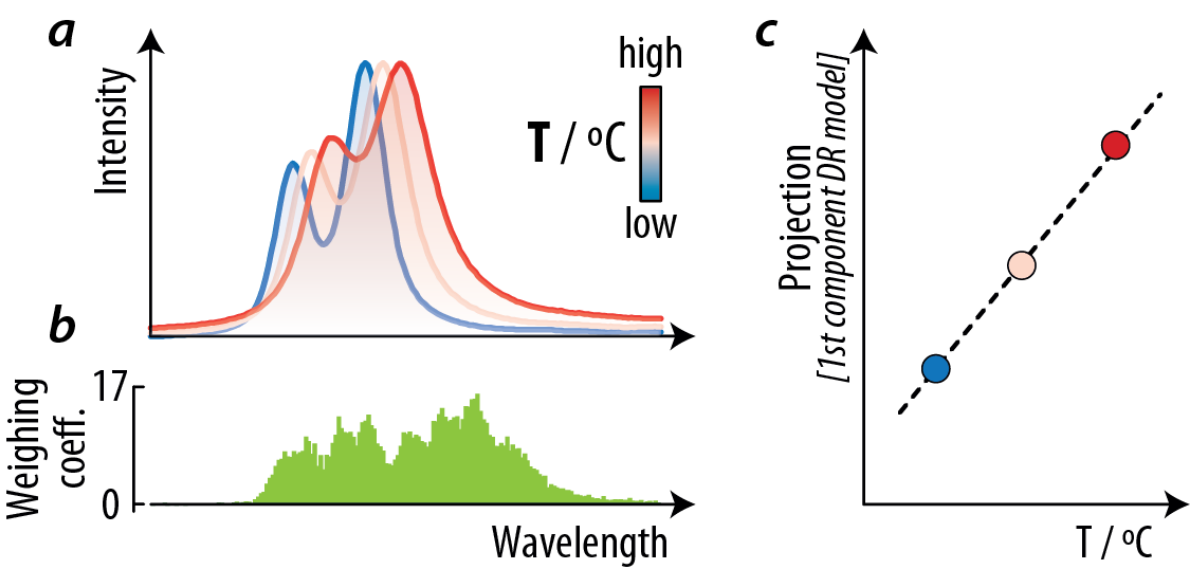

**Figure 3.** Graphical representation of a typical outcome of using DR algorithms (e.g., PCA, LDA) to analyse the calibration dataset. **a)** Three representative spectra of a calibration dataset are analysed via LDA to extract **b)** the weighing coefficients for the 1$^{st}$ component of the model. Multiplication of the (normalized) spectra by these coefficients afford a projection of the spectra that can be used as the readout feature **(c)** with the best trade-off in terms of specificity (accuracy) and sensitivity (precision).

## 2.2. Performance metrics

Several metrics have been introduced to quantify the performance of a luminescence thermometry approach and hence compare it with others based on different luminescent sensors and/or readout features. These are only briefly discussed below, and the interested reader is directed to relevant literature that describes them in detail.[13,41]

### 2.2.1. Precision

#### *i) Sensitivity*

The absolute thermal sensitivity[14,42,43] ($S_a$) represents the rate of change of a readout feature ($F$, e.g., position of the band maximum, bandwidth, intensity, polarization ratio, lifetime) as a function of temperature, as defined by equation 1. Its units will be nm/ºC for conformational changes, counts/ºC for intensity changes, and s/ºC for lifetimes measurements:

$$S_a = \frac{\partial F}{\partial T} \qquad \text{(eq. 1)}$$

However, comparing absolute sensitivities across thermometers using different readout modalities does not allow for a fair comparison of the performance. Therefore, the relative thermal sensitivity[44] ($S_r$, $^oC^{-1}$) is typically used as a "normalized" metric that levels differences that are readout modality-dependent, as shown in equation 2.

$$S_r = \frac{1}{F}\left|\frac{\partial F}{\partial T}\right| \qquad \text{(eq. 2)}$$

At variance with classical approaches (even more so in the case of readout features extracted from quantitatively known physical mechanisms), we note that the use of ML algorithms can lead to the identification of $F$ with values passing through 0 in the temperature calibration range. As such, the definition of $S_r$ becomes artificially high (tending to infinite) at $F = 0$. This is clearly a mathematical artifact. As such, a more meaningful metric to compare the performance of thermometric approaches is thermal uncertainty.

### *ii ) Uncertainty (or resolution)*

Temperature uncertainty ($\delta T$) represents the smallest temperature variation that can be reliably resolved by the thermometric approach. It depends on the intrinsic properties of the material and on the experimental setup. The value of $\delta T$ mirrors the statistical uncertainty in the determination of the readout feature, and is calculated as:[3,14]

$$\delta T = \pm\sqrt{\left(\frac{\partial T}{\partial F}\,\delta F\right)^2} = \pm\frac{1}{S_r}\left|\frac{\delta F}{F}\right| \qquad \text{(eq. 3)}$$

Hence, $\delta T$ is inversely proportional to the relative sensitivity ($S_r$). Note that in the case of purely intensity-based thermometric approaches $\delta F/F$ is nothing more than the signal-to-noise ratio (SNR, $\delta I/I$, where $I$ is the signal intensity), since $\delta F$ is the statistical fluctuation in the value of $F$ extracted from multiple acquisitions of the luminescence signal. Even though for other readout features the relation can become more complex, generally speaking the uncertainty is inversely proportional to the SNR value at which the calibration was carried out (!). As such, $\delta T$ should be treated as the lower limit of uncertainty achievable with a specific readout feature and modality. One cannot directly translate the value of uncertainty calculated from calibration datasets to experimental readouts, because in the latter situation SNR values are usually lower due to reduced probe concentration, low excitation intensities, and/or shorter integration times (all yielding lower overall signal).

Importantly, as per eq. 3, calculation of $\delta T$ does not strictly require knowledge of $S_r$. This is of utmost importance in the above-mentioned context of ML-driven thermometry, since $\delta T$ can be reliably used for quantifying the performance of the thermometric approaches.

### *iii) Noise Equivalent Sensitivity* [45]

Noise equivalent sensitivity (NES) is often overlooked by the community working in luminescence thermometry, as well as luminescence sensing more broadly speaking, including manometry, magnetometry, and velocimetry. This metric is possibly the most meaningful one (as also highlighted by Dionne and co-workers[45] since it factors in an additional aspect that the above parameters disregard. Specifically, NES can be thought of as the smallest detectable change appreciable by integrating over 1 s. As such, NES additionally "normalizes" the performance by integration time, reflecting the interplay between the detector sensitivity and the brightness of the luminescent probe.

Depending on the readout modality, NES is defined using different expressions, which are carefully derived by **Dionne** et al. considering a realistic situation of photon shot noise (Poissonian noise, see **Section 5**). The unit of NES is $°C/\sqrt{\mathrm{Hz}}$ (or $K/\sqrt{\mathrm{Hz}}$).

### 2.2.2. Accuracy

Accuracy (sometimes referred to as reliability) is quantified as the deviation between the experimentally measured temperature ($T_R$) and the reference (ground truth) temperature ($T_G$), typically obtained using a calibrated thermocouple, an infrared thermal camera, or theoretical models.

A smaller difference between $T_R$ and $T_G$ indicates higher accuracy. The mean absolute error (MAE; eq. 4) reflects the average magnitude of the errors. On the other hand, the root mean square of the residuals (RMSE), eq. 5, can also be used and it strongly penalizes larger deviations, being more sensitive to outliers or occasional large deviations.

$$MAE = \left|\frac{\Sigma_i \ (T_{R,i} - T_{G,i})}{n}\right| \qquad \text{(eq. 4)}$$

$$RMSE = \frac{\Sigma_i \sqrt{(T_{R,i} - T_{G,i})^2}}{n} \qquad \text{(eq. 5)}$$

Here the subscript *i* indicates the *i*-th measurement.

### *i) Cross-sensitivity*

Cross-sensitivity is the effect of non-target (interfering) parameters on *F* through direct interaction with the sensor, which results in a confounding effect on the readout of the target parameter (temperature in this context). Cross-sensitivity is a major limitation in intracellular thermometry,[3,46] since the cell microenvironment is highly dynamic and heterogeneous, characterized by variations in the value of physicochemical parameters such as pH, viscosity, refractive index, and ionic strength. The possible modulation of *F* by these parameters thus reduces the accuracy of temperature determination in an unpredictable manner.

Therefore, minimization of the impact of cross-sensitivity requires improving the specificity of the sensor via appropriate design. To assess and quantify the effect of cross-sensitivity, calibration measurements should be carried out under strictly controlled conditions to explore the multidimensional space described possible interfering parameters. Nature and range of these parameters should be selected carefully based on the final use of the thermometer. Recently, by building on the temperature (pressure)-invariant manometric (thermometric) factor (TIMF and PITF) proposed by Marciniak and co-workers, we proposed a unifying working metric that allows quantifying loss of accuracy in selected ranges of target and non-target parameters.[32] Cross-sensitivity tolerance (CST) is defined as the ratio between the absolute sensitivities of non-target (*X*) and target (*T*, for temperature) parameter:

$$CST = \frac{\partial F}{\partial X} \Big/ \frac{\partial F}{\partial T} = \frac{S_{a,X}}{S_{a,T}} \qquad \text{(eq. 6)}$$

CST can then be used to build a relative error map (**Figure 4**) that shows the relative loss of accuracy ($\frac{\Delta T}{T}$) induced in the readout of temperature when the non-target parameter changes by a certain amount ($\Delta X$).

$$\frac{\Delta T}{T}(\%) = \left|\frac{\Delta X}{X \cdot CST}\right| \cdot 100 \qquad \text{(eq. 7)}$$

### *ii) Bias*

Bias refers to the loss of accuracy caused by measurement conditions and procedures.[25,47] One can visualize this as a process that affects the luminescence signal after the photons have left the probe. For instance, biological media can introduce spectral distortions due to wavelength-dependent photon absorption.[47] The same is true for wavelength-dependent response of optical elements in the optical path. Bias is generally observed as a systematic error in the thermal readout.

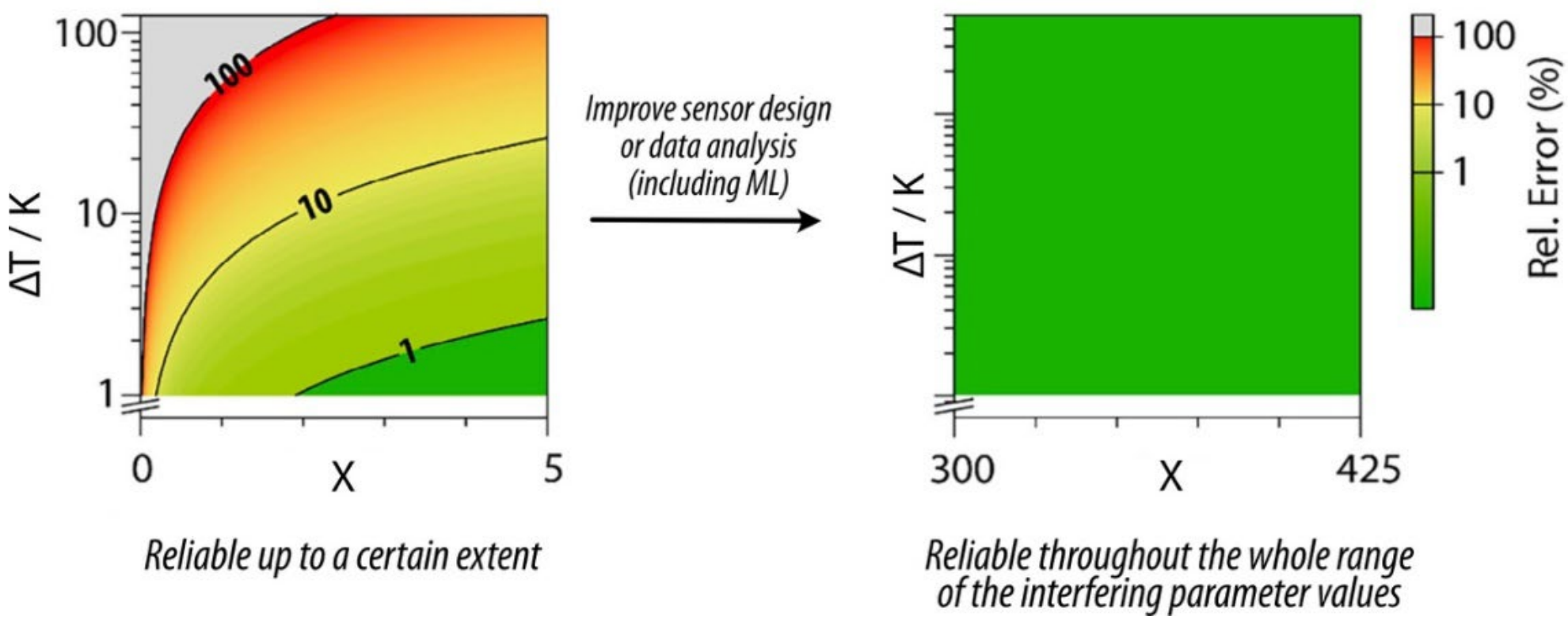

**Figure 4.** Quantification of the relative error in the target parameter readout as a function of the interfering parameter (X). 1%, 10%, and 100% relative error thresholds are indicated to exemplify the delineation of the operable-inoperable X–T space regions based on the permittable error threshold.

***iii) Repeatability***

Repeatability describes the ability of the thermometer to consistently produce the same result under identical conditions. A repeatability above 90% is generally considered satisfactory.[42] It is defined by equation 8:

$$R = 1 - \frac{\max(|F_i - F_{mean}|)}{F_{mean}} \quad \text{(eq. 8)}$$

where $F_i$ is the readout at a given temperature for the $i$-th measurement, and $F_{mean}$ is the average value obtained from $n$ independent measurements. Repeatability is generally dictated by the quality of the probe, which should be resistant against temperature-induced irreversible changes (e.g., heat-induced trap state formation, surface ligand removal, and thermal degradation) and other phenomena like photobleaching. Ultimately, repeatability is chiefly affected by the tendency of the probe to show hysteresis.

### 2.2.3. Reproducibility

Reproducibility refers to the consistency of measurements obtained under different conditions, such as varying experimental methods, equipment, operators, or time periods.[14] This is one of the most underexplored aspects, which, nonetheless, is crucial to move towards standardization of the field. Cross-laboratory tests are required in this sense, involving metrology institutes alongside research institutions. Notably, the interest in this type of activity is very much alive at the European level, as underscored by the Euramet project LUMETH, in which a metrological framework for the reliable use of luminescence thermometry in healthcare between top level research groups working on this field.[48]

### 2.2.4. Spatial and Temporal Resolution

Spatial and temporal resolutions are defined as the minimum spatial distance or time interval between two measurements that exhibit a detectable temperature difference larger than the thermal resolution ($\delta T$).[15,35] In other words, spatial resolution determines how close two regions can be while still being distinguishable as having different temperatures, whereas temporal resolution describes the shortest time scale over which temperature variations can be reliably detected. Both parameters are ultimately limited by the sensitivity of the thermometric probe, the SNR, and the acquisition capabilities of the detection system.[14]

In intracellular thermometry, spatial resolution is particularly relevant since this parameter is intrinsically limited by optics. Specifically, unless super-resolution approaches are used,[49–51] for diffraction-limited measurements the values of lateral ($d_{xy}$) and axial ($d_z$) spatial resolutions are determined by the working wavelength, the characteristics of the objective, and the refractive index of the immersion medium. Further details about spatial resolution for different microscopy approaches are given in **Section 4**.

## 2.3. Thermal physics at the cellular scale

Having defined the main readouts and performance metrics governing luminescent thermometry, it is essential to consider the physical limits that determine whether the measured intracellular temperature variations are physically plausible. In principle, if cross-sensitivity effects and experimental biases are properly controlled, experimental observations should be consistent with theoretical predictions based on heat transfer and diffusion at the microscale. Therefore, apparent discrepancies between measured and predicted temperature gradients (**Figure 5**) may either reveal previously unaccounted (or inappropriately accounted) biophysical processes or suggest the presence of uncontrolled experimental artifacts. Understanding the fundamental principles of heat transfer and diffusion is thus crucial for correctly interpreting intracellular thermal measurements and assessing their biological relevance.[52]

### 2.3.1. Heat transfer: Fourier's law

At the cellular level, heat transfer can be first approached using Fourier's law, which relates the local heat flux to the temperature gradient. Baffou and co-workers[27] reformulated this relation for steady-state conditions as:

$$\Delta T = \frac{P}{\kappa L} \quad \text{(eq. 9)}$$

where $\Delta T$ is the temperature increase generated by a heat source of power $P$, $\kappa$ is the thermal conductivity of the medium, and $L$ is the characteristic size of the heat source—even though Kiyonaka et al. interpret it as the distance from the centre of the heat source.[3,27,43] For a cell, Baffou identified typical values of $P$, $\kappa$, and $L$ to be $10^{-12}$–$10^{-11}$ W, 0.4–0.6 W $m^{-1}$ $K^{-1}$, and 10 µm, yielding $\Delta T$ values ≤ $10^{-4}$ K. Even considering the smaller size of mitochondria (≈ 0.5 µm) temperature increments of > 1 K at the organelle's surface are not realistically achievable. The discrepancy between temperature values obtained from simulations and experimentally should, thus, originate from effects related to cross-sensitivity or bias.

While this interpretation is often accepted, some researchers (Uchiyama, Okabe, Jaque, Brites, among others[28,30,31]) argue that the simplified model neglects the heterogeneous composition of the cytoplasm, interfacial thermal resistance, and transient power spikes occurring during metabolic or photothermal processes.

### 2.3.2. How fast heat spreads: Diffusion

The characteristic time for a thermal disturbance to spread over a given length scale through heat diffusion can be approximated as:[27,53]

$$\tau_{th} \sim \frac{d^2}{\alpha} \quad \text{(eq. 10)}$$

Where

$$\alpha = \frac{k}{\rho c_p} \quad \text{(eq. 11)}$$

$d$ is relevant length scale over which heat diffuses, $k$ is the thermal conductivity, $\rho$ is the medium density, $c_p$ is the heat capacity, and $\alpha$ (in units of $\mathrm{m}^2\ \mathrm{s}^{-1}$) is the thermal diffusion constant, which determines the rate at which heat propagates through the medium. For characteristic cellular dimensions ($d$ ≈ 5–10 μm), and considering a value $\alpha \approx 1.4 \cdot 10^{-7} m^2 s^{-1}$ (water), the thermal diffusion time is in the millisecond scale at best, a value supported by both simple analytical and numerical models[27,54] as well as bulk[55] microcalorimetric data. This extremely rapid diffusion indicates that heat equilibrates almost instantaneously relative to biological timescales.

### 2.3.3. Interfacial thermal resistance

Temperature discontinuities can arise at boundaries due to Kapitza resistance—defined as the resistance to heat flow that occurs at the interface between the materials.[56] When this resistance is significant, the temperature field shows clear discontinuities, which can dominate heat transport, modify stresses, reduce thermoelectric efficiency, and affect fluid or electronic behaviour at micro- and nanoscale interfaces. Note that the presence of a measurable effect of the Kapitza resistance would also have implication on the capability of a luminescent thermometer trapped, e.g., in an endosome to accurately pick up temperature variations within the cell. Again, Baffou notes that thermal discontinuity at a biological membrane cannot exceed $10^{-5}$ K, casting doubts on the validity of the experimental observations of higher temperatures in the cell nucleus or mitochondria compared to the surroundings.[10,29,57–61]

## 3. Considerations on thermometer design

As explained above, a luminescent thermometer is a probe that responds to temperature variations through measurable changes in its emission properties.[62] Luminescent thermometers include small dye molecules, metal complexes, luminescent micelles, as well as organic and inorganic nanoparticles, such as fluorescent polymeric dots, carbon dots, quantum dots, and lanthanide-doped nanoparticles.[63–68] Species that have been implemented as luminescent thermometers for intracellular sensing include inorganic nanoparticles (lanthanide-doped nanoparticles, quantum dots, carbon dots, nanodiamonds), lanthanide-micelles, metal complexes, organic dyes, modified biomolecules (proteins above all), and polymeric probes.[9,69–87]

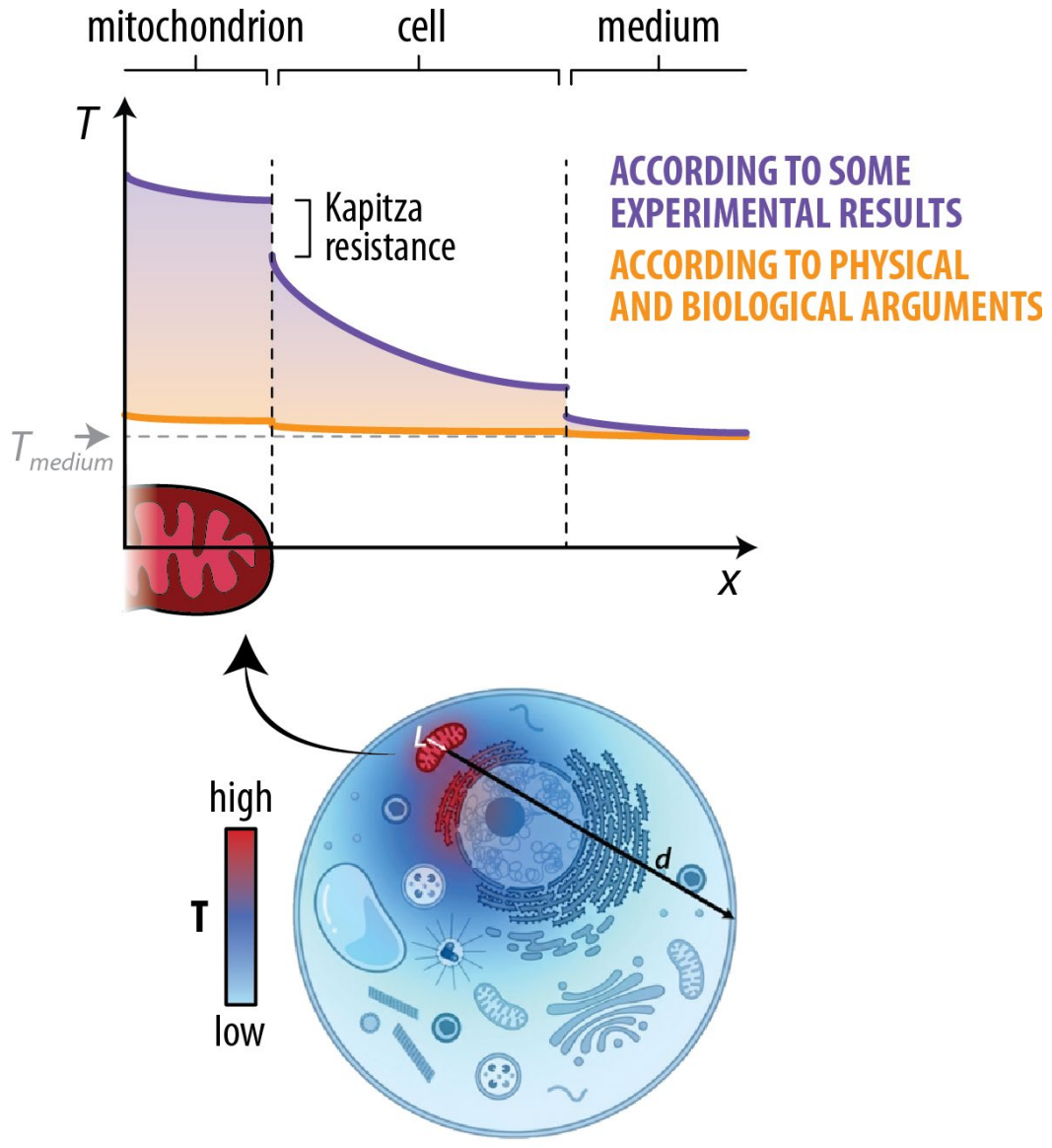

**Figure 5.** Illustration of thermal physical parameters involved in a cell. Expected trends according to physical and biological arguments (orange) and according to some experimental observations (purple).

Regardless of the type of the species selected, a luminescent thermometer should meet several criteria to achieve accurate intracellular temperature mapping (see **Table 2**): a working range encompassing the temperature values to be probed, responsiveness in a timescale shorter than the one of the events to be probed, high brightness, photostability, and minimal cross-sensitivity to environmental factors such as pH, ionic strength, or biomacromolecules.[75,3] This last requirement is crucial—and possibly the most difficult to meet—for reliable intracellular thermometry, and it should be ensured by-design rather than verified a posteriori. Some approaches to mitigate cross-sensitivity include preparation of core/shell architectures and maximization of the hydrophobicity of the compartment in which the temperature-sensitive part of the probe is located.[88–90] Chiefly, the properties of the thermometer should be consistent from batch to batch, whether produced in a single or across different laboratories, hence setting the basis for cross-validation and broader implementation.

The chemistry of the thermometers should also be finely controlled considering the interaction with the cellular environment. More specifically, the thermometers should show negligible cytotoxicity at the concentration that will be subsequently used for cell studies. In the case of inorganic nanoparticles, materials with low solubility constant (such as $Ag_2S$[91–95]) are to be preferred to minimize cation release.

Not less importantly, the thermometers should be taken up by the cells and, depending on the study to be carried out, they should accumulate in specific cell compartments. There are numerous strategies for targeting specific organelles, and they mainly rely on the use of organelle-targeting peptides or the introduction of specific functional groups.[96,97] Of particular interest in the context of luminescence thermometry are the approaches for targeting mitochondria.[98] Triphenyl phosphonium (TPP) is a functional group with known mitochondria-targeting capabilities, while dyes with cationic and aromatic nature[99] (e.g., rhodamines, rosamines, cyanines) often naturally accumulate in these organelles.

Targeting specific organelles is further complicated by the need to achieve endosomal escape. While endosomal entrapment occurs also for molecules, this process is exacerbated by the size and often heavily charged nature of nanoparticles.[100] Investigation on this subject is spurred by therapy development, with the most successful strategies—albeit not the only ones[101]—relying on the acidic pH of endosomes/lysosomes (which progressively decreases from 6.5 to 4.5 during the full endosome-to-lysosome cycle).[100] These approaches include leveraging the proton sponge effect and the use of pH-sensitive polymers and peptides.

When a new luminescent thermometer is developed, the internalization and accumulation in the organelle/compartment of interest should be verified. The simplest and most effective approach is performing a colocalization experiment, whereby the spatial distribution of the probe's luminescence in a fluorescence microscope image is compared with the ones of fluorescent species with known organelle-targeting capabilities.[102]

Ultimately, the development of reliable luminescent molecular thermometers requires a rigorous and systematic assessment of all photophysical, physicochemical, and environmental parameters that can influence their thermal response. An overview of the biological and photophysical considerations[1,3,31] to be taken into account when developing/choosing a luminescent thermometer for intracellular thermometry are listed in **Table 2**.

# 4. Instrumentation setup and environmental control

In this section we describe the most common instrumentation used for intracellular luminescence thermometry. Most experimental implementations rely on a single core instrument: an inverted fluorescence microscope. This microscope constitutes the central body of the setup and serves as a modular foundation that can be adapted to different imaging modalities depending on the experimental requirements (**Figure 6**).

## 4.1. Fluorescence microscopy as a modular platform

In all cases, luminescent probes within the sample are excited using light of an appropriate wavelength and emit photons at longer (downshift)[103] or shorted (photon upconversion, including avalanching, or two-photon excitation) wavelengths.[104,105] These emitted photons are then separated from the excitation light by means of dichroic mirrors and optical filters and subsequently detected using cameras, photomultipliers, or spectrometers (**Figure 6**).

Importantly, the same inverted microscope platform can be reconfigured with different optical paths, excitation sources, and detection units to implement distinct measurement modalities.

**Table 2.** Essential photophysical properties, optical and biological considerations for designing thermometers.

| PROPERTY | DEFINITION / ADVANTAGES | DOWNSIDES / CAVEATS | BIO AND TECHNICAL RELEVANCE |
|---|---|---|---|
| High brightness (B) | Defined as the product of extinction coefficient (ε) and emission quantum yield (ϕ). | Achieving high brightness often requires complex structural engineering, larger probe sizes, or heavier metal complexes which can impact solubility, cellular delivery, or cytotoxicity. | Enables detection with standard, cheap photon collection tech; allows low probe concentration, short integration times (high temporal resolution/SNR), and controlled excitation intensity to avoid laser-induced sample heating. |
| Photostability | Ability to sustain prolonged illumination without significant photobleaching or photodecomposition. | Inorganic probes are often more photostable than organic ones, but can suffer from slow clearance rates, potential long-term toxicity, or complex surface functionalization requirements. | Essential for continuous or long-term tracking, though the exact duration requirement varies depending on the specific measurement type or study being performed. |
| Spectral Integrity | Narrow and well-separated emission bands that minimize spectral overlap. | Traditional organic fluorophores typically exhibit broad, asymmetrical emission bands with prominent side-bands, naturally limiting multi-color multiplexing unless complex deconvolution is used. | Facilitates clean readout analyses in classical luminescence thermometry. Note: This requirement is starting to be alleviated by emerging machine learning (ML)-assisted spectral analysis. |
| Minimal self-absorption | Preventing overlap between the absorption and emission spectra to avoid spectral distortions from signal reabsorption. | Probes with minimal (pseudo) Stokes shifts suffer heavily from this effect; correcting it requires more complex probe design and rigorous validation to ensure concentration-dependent artifacts are absent. | Requires meticulous cell-washing procedures during sample preparation so that only probes incorporated into cells or attached to membranes remain. |
| Emission range: Visible | Generally brighter and more stable (especially organic dyes lacking highly conjugated unsaturated structures). Detectors are highly sensitive and cheap. | Potential absorption from dyes in the culture media, high cellular autofluorescence, and frequently requires phototoxic UV excitation. | Best suited for high-signal applications where background autofluorescence can be managed or gated, and budget is a constraint. |
| Emission range: NIR | Highly reduced scattering and negligible biological autofluorescence. Does not require UV excitation. | Often less bright (lower PLQY) and less photostable (especially dyes). NIR cameras are less sensitive, noisier, and 1 to 2 orders of magnitude more expensive. | Ideal for deep tissue or complex cellular environments where minimizing background autofluorescence is critical, despite higher hardware costs. |
| Biocompatibility | The requirement that the probe does not interfere with cellular metabolism. | Strict biocompatibility criteria drastically narrow down the chemical library of usable materials, often forcing a compromise on brightness, photostability, or target specificity to avoid cytotoxicity. | Confirms the thermometer does not interfere with the physiology of the biological system/environment. |

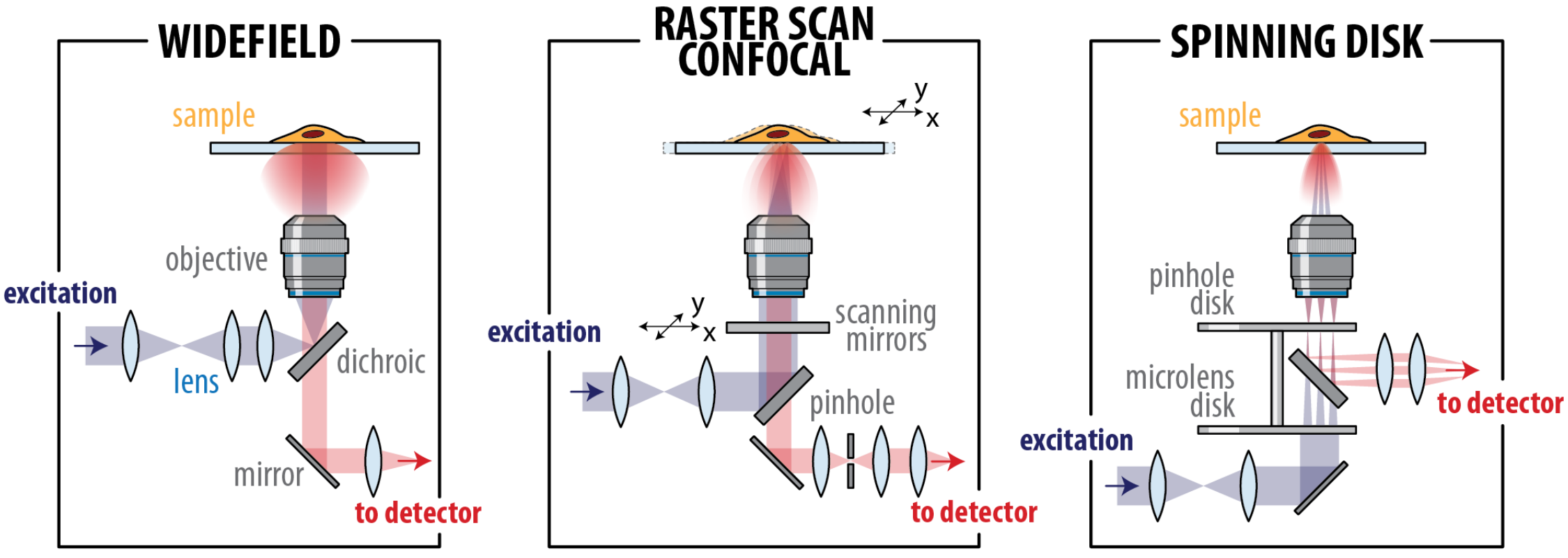

**Figure 6.** Representation of the main components of the three possible architectures described in this Section.

These must be clearly distinguished, as they correspond to fundamentally different readout strategies: i) Intensity-based imaging, including widefield and confocal imaging, yields spatially resolved intensity maps, $I(x, y)$; ii) Spectral imaging records spectrally resolved information, generating datasets of the form $I(x, y, \lambda)$, including multispectral and hyperspectral imaging systems; iii) Fluorescence lifetime imaging microscope (FLIM) records the temporal evolution of the luminescence signal, $I(x, y, t)$, from which fluorescence lifetime maps $\tau(x, y)$ can be extracted. Each modality makes use of specific combinations of optical components and detection schemes integrated into the same microscope platform, enabling access to complementary intensity-, spectra-, or lifetime-based thermometric information (**Figure 6**).

## 4.2. Intensity-based imaging modalities

Intensity-based imaging is the simplest and most widely used implementation of fluorescence microscopy. In this approach, the emitted photons are detected by imaging sensors to produce spatial maps of fluorescence intensity. Depending on the excitation and detection geometry, two main configurations are commonly employed: widefield and confocal microscopy.

### 4.2.1. Widefield fluorescence microscopy

When the microscope is coupled to imaging cameras (e.g., (EM)CCD or (s)CMOS detectors; see below), fluorescence signals are recorded simultaneously from the entire field of view (FoV). Each pixel of the detector measures the local fluorescence intensity, generating an image $I(x, y)$ that reflects the spatial distribution of the luminescent signals within the sample. The brightness of each pixel reflects the local emission from the luminescent probes (and/or endogenous dyes) and therefore reports their spatial distribution within the sample.

Widefield imaging offers high acquisition speed and excellent photon collection efficiency, making it particularly suitable for dynamic processes and large-area imaging. However, because fluorescence originating from different focal planes is collected simultaneously, out-of-focus background signals can reduce image contrast and limit axial resolution, especially in thick biological specimens.

### 4.2.2. Confocal fluorescence microscopy

Confocal microscopy overcomes many of the limitations of widefield imaging by restricting both excitation and detection to a diffraction-limited focal volume. While in widefield microscopy the excitation is distributed across the entire FoV, in confocal microscopy the excitation is concentrated at a specific point of the specimen. The waist of the focused excitation beam defines the area that is illuminated at any given time. Fluorescence emitted from this focal volume is collected through the same objective and directed towards the detection system.

The term "confocal" originates from the use of pinholes positioned at conjugate focal planes along the optical path. These pinholes reject out-of-focus emission originating from regions above and below the focal plane, thereby improving image contrast and axial resolution. As a result, confocal microscopy enables optical sectioning and the acquisition of three-dimensional image stacks (Z-scans) via mechanical scan of the sample along the vertical direction, making it particularly valuable for intracellular studies and volumetric imaging.

Several implementations of confocal microscopy are commonly employed. In conventional **laser-scanning confocal microscopy (LSCM)**,[106,107] a focused laser beam is raster-scanned (*xy*) across the specimen using galvanometric mirrors, and the image is reconstructed point-by-point from the detected fluorescence signal. Alternatively, the sample can be moved in controlled steps. This approach provides flexibility in excitation and detection schemes, although image acquisition can be relatively slow due to the sequential nature of the scanning process.

An alternative implementation is **spinning-disk confocal microscopy**,[108] which employs a pair of synchronized Nipkow disks rotating rapidly and containing thousands of pinholes or microlenses. Instead of illuminating a single point at a time, multiple excitation spots are generated simultaneously across the field of view, allowing parallel acquisition of fluorescence signals. This configuration significantly increases imaging speed while maintaining optical sectioning capabilities, making it particularly attractive for live-cell imaging and fast biological processes. However, compared with laser-scanning systems, spinning-disk microscopes may offer lower spectral flexibility and reduced photon collection efficiency in some applications, while also being generally more expensive.

Other variants, such as resonant-scanning and line-scanning confocal microscopes, have also been developed to increase imaging speed, although laser-scanning and spinning-disk systems remain the most widely-used configurations in intracellular luminescence thermometry.[109]

## 4.3. Spectral imaging modalities

While intensity imaging provides information on the spatial distribution of luminescent probes, many luminescent thermometers rely on temperature-dependent spectral features such as emission intensity ratios, peak shifts, or bandwidth variations. Consequently, the detection system must be capable of recording spectral information.

Both widefield and confocal microscopes can therefore be equipped with spectral detection modules, enabling the acquisition of spectral information with different levels of spatial and spectral resolution. Depending on the balance required between spectral content, spatial information, and acquisition speed, three main approaches can be distinguished: spectral acquisition, multispectral imaging, and hyperspectral imaging.

***i) Spectral acquisition.*** The simplest implementation consists of coupling the microscope to a spectrometer. In this configuration, the emitted light is dispersed according to wavelength and recorded as an emission spectrum, yielding an intensity-vs-wavelength dataset, $I(\lambda)$. Since the detected signal corresponds to the averaged emission originating from the observed region, spatial information is sacrificed in favour of spectral resolution. This approach enables the extraction of temperature-dependent spectral parameters such as peak shifts, bandwidth variations, or intensity ratios between different emission bands. Yet, it gives a very coarse-grained view of the thermal state of the specimen, since it averages the spectral response over the whole FoV or selected region of interest (ROI). This approach can be used during the calibration step to acquire the whole spectral information before implementing multispectral acquisition.

***ii) Multispectral acquisition.*** An intermediate approach consists of acquiring images in several predefined wavelength ranges while preserving spatial information, a strategy commonly referred to as multispectral imaging. This can be achieved either sequentially, using filter wheels or tuneable filters, or simultaneously, using beam splitters and multiple detectors. In both cases, images are collected at different spectral bands and combined into a stack of spectral channels, yielding a dataset of the form $I(x, y, \lambda_i)$—where $\lambda_i$ is the center wavelength of the *i*-th range.

If a sufficiently large number of narrow bandpass filters is employed, the resulting image stack can be used to reconstruct approximate spectral profiles at the pixel level and, in some cases, generate hyperspectral-like datasets, although with lower wavelength resolution than that achievable with dedicated hyperspectral systems. An alternative implementation consists of simultaneously projecting signals filtered through different spectral channels onto the area of a single sensor.[110–112] This approach can significantly improve temporal resolution—since all the projections are imaged simultaneously on the same sensor area—albeit reducing the FoV size.

Multispectral acquisition therefore occupies an intermediate position between conventional intensity imaging and hyperspectral imaging, providing a practical compromise between spectral information content, acquisition speed, data volume, and experimental complexity. It should also be noted that the definition of multispectral imaging is relatively loose, since any acquisition involving two or more spectral channels can formally be considered multispectral. In this sense, even standard RGB cameras may be regarded as simple multispectral detectors. Moreover, if enough spectral bands are used, one is effectively moving towards hyperspectral acquisitions.

***iii) Hyperspectral acquisition.*** More advanced implementations extend this concept to spatially resolved spectroscopy through hyperspectral imaging. This approach generates a hyperspectral data cube $I(x, y, \lambda)$, which contains simultaneously spatial and spectral information. Thus, this approach enables the extraction of spectral readout features at the single-pixel level and the reconstruction of high-resolution temperature maps.

Historically, hyperspectral imaging has often been implemented using a confocal microscope, where a full emission spectrum is acquired at each position during the raster-scanning process.

The augmented spectral information with (sub-)micrometric spatial resolution comes at a cost of the abovementioned reduced temporal resolution generally characterizing confocal microscopy.

To overcome this limitation, several widefield implementations have been developed. These include systems based on wavelength-selective optical elements placed in front of the detector, such as Fabry–Pérot interferometers, acousto-optic tuneable filters, or liquid crystal-based tuneable filters, which enable the acquisition of image stacks at different wavelength bands. In some cases, mosaic or snapshot hyperspectral cameras are used to directly capture spectrally resolved images without scanning. These approaches can substantially increase acquisition speed, although they often involve trade-offs in spectral resolution, wavelength coverage, spatial sampling, or instrument cost.

In three-dimensional implementations, optical sectioning can be combined with multispectral or hyperspectral detection to generate datasets of the form $I(x, y, z, \lambda)$, enabling volumetric spectral mapping.

Ultimately, the choice between spectral, multispectral, and hyperspectral acquisition reflects a compromise between spectral resolution, spatial resolution, temporal resolution, data complexity, and instrument cost.

## 4.4. Lifetime detection and fluorescence lifetime imaging microscopy (FLIM)

In contrast to intensity- and spectra-based measurements, fluorescence lifetime measurements exploit the temporal evolution of the luminescence signal following excitation.[94] Consequently, the recorded signal is the temporal decay profile, ($I(t)$), from which the fluorescence or luminescence lifetime ($\tau$) can be extracted.[113,114]

Fluorescence lifetime measurements can be implemented by coupling the microscope to fast point photodetectors, such as avalanche photodiodes (APDs)[115,116] and photomultiplier tubes (PMTs),[117,118] together with time-resolved acquisition electronics (e.g., time-correlated single-photon counting, TCSPC).[119,120] In this configuration, the temporal decay of the fluorescence signal following pulsed excitation is recorded to determine the fluorescence lifetime. TCSPC is generally regarded as the gold standard for lifetime determination owing to its excellent temporal resolution and accuracy. Alternative implementations rely on time-gated detection schemes, in which the emitted signal is integrated within predefined temporal windows using gated cameras. Both approaches ultimately provide access to the fluorescence lifetime, albeit with different compromises in temporal resolution, acquisition speed, and experimental complexity.

When lifetime measurements are performed across the entire FoV, this approach becomes fluorescence lifetime imaging microscopy (FLIM),[121,122] which generates spatial maps of lifetimes. Several FLIM implementations are available, differing mainly in the microscope architecture and detection strategy employed.

In laser-scanning (or sample-scanning) FLIM,[123,124] lifetime information is acquired sequentially during the raster-scanning process, typically using confocal microscopes coupled to TCSPC electronics. This configuration provides excellent spatial resolution and accurate lifetime determination, making it one of the most widely-used implementations for intracellular thermometry. However, the sequential acquisition process can limit imaging speed, particularly when large FoVs or high photon counts are required.

Widefield FLIM,[125,126] by contrast, records lifetime information across the entire field of view simultaneously. This modality is commonly implemented using time-gated cameras, such as gated EMCCD or sCMOS detectors, allowing for rapid acquisition of lifetime maps. Owing to its high temporal resolution and efficient use of emitted photons, widefield FLIM is particularly attractive for monitoring dynamic biological processes. However, as in conventional widefield microscopy, the lack of optical sectioning can reduce image contrast in thick specimens.

Intermediate approaches based on spinning-disk confocal microscopy[127] combine optical sectioning with parallelized image acquisition. By simultaneously illuminating multiple points through rotating Nipkow disks, spinning-disk FLIM can achieve substantially higher imaging speeds than conventional laser-scanning FLIM while maintaining confocal sectioning capabilities. This makes it particularly suitable for live-cell studies,[128] where cellular motion and rapid physiological dynamics may compromise slower acquisition methods. Nevertheless, the implementation of lifetime detection in spinning-disk systems is generally more complex and remains less widespread than conventional laser-scanning FLIM.

While the term FLIM remains mainstream, it carries the historical legacy of using fast-emitting fluorophores—where fluorescence dominates—in conventional fluorescence microscopy. As such, FLIM is mainly carried out with fast emitters (ns timescale). Modern setups, however, are increasingly becoming capable of measuring lifetimes in the micro- and millisecond range, which are characteristic of, e.g., phosphorescence or delayed emission from lanthanides.

Overall, fluorescence microscopy platforms equipped with different detection modules provide a versatile toolbox for intracellular thermometry, enabling temperature measurements based on intensity, spectral features, or fluorescence lifetime.

In practice, the choice of microscope architecture involves trade-offs between spatial resolution, optical sectioning capability, acquisition speed, experimental complexity, and cost. These factors ultimately determine the suitability of a given system for a particular thermometric application. A comparison of the most common microscope architectures used in intracellular luminescence thermometry is provided in **Table 3**.

**Table 3.** Comparison of microscope architectures commonly used for intracellular luminescence thermometry.

| | WIDEFIELD | LASER-SCANNING CONFOCAL | SPINNING-DISK CONFOCAL |
|---|---|---|---|
| OPTICAL SECTIONING | No | Yes | Yes |
| AXIAL RESOLUTION | Low | High | High |
| LATERAL RESOLUTION | Moderate | High | High |
| ACQUISITION SPEED | Very high | Low-Moderate | High |
| LIVE-CELL COMPATIBILITY | Good | Moderate | Excellent |
| PHOTOBLEACHING | Low | Moderate-High | Moderate |
| PHOTON COLLECTION EFFICIENCY | High | Moderate | High |
| SPECTRAL IMAGING COMPATIBILITY | Good | Excellent | Moderate |
| FLIM COMPATIBILITY | Good (time-gated) | Excellent (TCSPC) | Good |
| EXPERIMENTAL COMPLEXITY | Low | High | High |
| COST | Low | High | High |
| TYPICAL THERMOMETRY APPLICATION | Population screening | High-resolution intracellular thermometry | Dynamic intracellular thermometry |

### 4.4.1. Detection system selection (Noise level, Dynamic range, Quantum efficiency, Pixel size, Shutter modality)

Detection systems used in microscopy mainly include area detectors (CCDs, EMCCDs, CMOS, sCMOS) and point detectors (photomultipliers, PMTs and (avalanche) photodiodes, APDs; **Figure 7**).

The building blocks of all area detectors (cameras) are photodiodes (PDs): solid-state, semiconductor-based point detectors. In a PD, the impinging photon produces electron-hole pairs that are accelerated at the opposing electrodes of the diode, producing a measurable current, which is proportional to the number of photons. Depending on the readout modality of the charge generated at each PD (pixel) making up the sensor surface, two main families of cameras are identifiable: CMOSs, where the signal is processed directly at each individual pixel, and CCDs, where the charges in each pixel raw are transferred to register rows to be read in batches. Without entering into the technical details of these two sensor families, they are both used for two-dimensional imaging as well as for spectral acquisition (when coupled with spectral dispersing elements).

In the family of point detectors, besides PDs, we find APDs and PMTs (**Figure 7**). APDs are akin to PDs, but they have a slightly more complex architecture and work at higher voltages (close to the electrical breakdown voltage) to ensure signal amplification. The strong applied electric field greatly accelerates the carrier inducing multiple impacts with the lattice, which effectively yield an avalanche of charge carriers. PMTs are instead vacuum phototubes that operate through the photoelectric effect: Incident photons generate photoelectrons at a photocathode, which are subsequently amplified through a cascade of dynodes in a vacuum tube to produce a measurable electrical signal. APDs and PMTs offer unrivalled speed and sensitivity, being used mainly for time-resolved photoluminescence studies, including FLIM but also for raster scan confocal imaging.

Below, we offer a description of the practical considerations to be considered when operating both families of detectors.

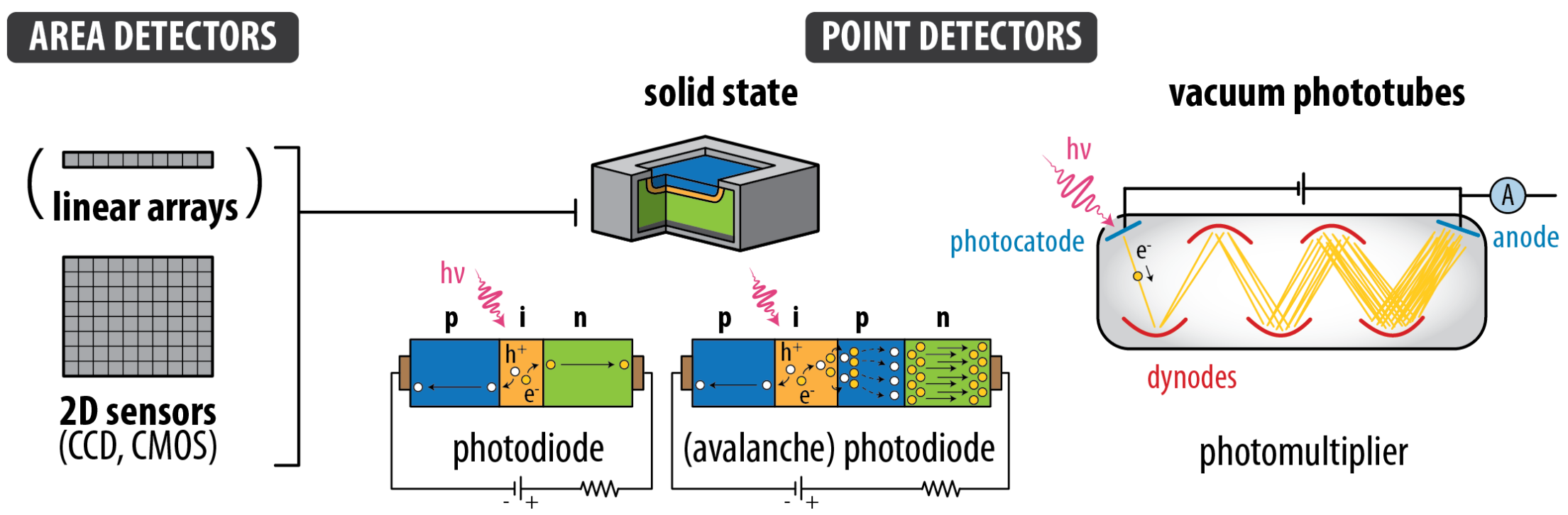

**Figure 7.** Types of detectors generally used in different microscopy modalities employed in intracellular luminescence thermometry.

**Area detectors.** The total noise for a solid-state detection system (**Figure 8**) is primarily composed of read noise ($\sigma_R$, constant per pixel), together with dark and photon shot noises ($\sigma_D$ and $\sigma_S$, both scaling with the square root of the exposure time, given the Poisson nature of the particle distributions).

$$\sigma_{eff} = \sqrt{\sigma_D^2 + \sigma_R^2 + \sigma_S^2} \qquad \text{(eq. 12)}$$

where read noise ($\sigma_R$) is a parameter related to the conversion of photons into intensity values through electronics. Dark noise ($\sigma_D$) is related to the dark current ($I_D$, electrons thermally generated at the pixel without illumination), while photon shot noise ($\sigma_S$) stems from the random fluctuations in the incoming discrete photons (total signal $S$). Both dark and signal electrons follow a Poisson statistic. As such dark noise is given by:

$$\sigma_D = \sqrt{I_D t} \qquad \text{(eq. 13)}$$

Where $I_D$ is the dark current and $t$ is the exposure time. The total signal ($S$) can be expressed as:

$$S = (QE)Nt \qquad \text{(eq. 14)}$$

where $QE$ is the quantum efficiency of the detector and $N$ is the rate of incident photons per second. Consequently, the shot noise is given by:

$$\sigma_S = \sqrt{(QE)Nt} \qquad \text{(eq. 15)}$$

$QE$ represents the probability that an incoming photon generates a measurable electron within the detector. High $QE$ values directly increase the detected signal and improve the achievable SNR, particularly in photon-limited experiments (i.e., when the system is photon starving). Consequently, $QE$ is a critical parameter when selecting detectors for intracellular thermometry, especially when weak luminescence signals are expected.
Total noise affects SNR values and background levels and is highly dependent on the type of detector/camera selected ($\sigma_R$) as well as the experimental conditions (above all, the acquisition time, for shot noises). $\sigma_R$ tend to contribute at short exposure times, while $\sigma_D$ and $\sigma_S$ take over at longer exposure times (when large numbers of photons are collected). Cooling of the sensor reduces $\sigma_D$, yet it also affects the operation range and spectral response of the sensor via modulation of the *QE* curve. Because of this latter effect, the use of consistent detector cooling is critical for reliable thermal mapping when using some spectral readout modalities (e.g., LIR). Typical $\sigma_R$ and $\sigma_D$ values reported for the detector technologies most commonly employed in intracellular luminescence thermometry are summarized in **Table 4**.[129–138]

To put the noise level into context, a more complete view is offered by the maximum SNR, which, in turn, is bound by the square root of the full well capacity (FWC). The FWC is the number of electrons that can be accumulated at a pixel before it saturates. This is an important parameter especially for sensing applications. Higher values of FWC ensure that a wider range of intensities can be probed before saturation kicks in while also generally ensuring that linearity in the response of the sensor is retained for different exposure times. This is pivotal in spectra-based thermometry modalities, which are affected by non-linear sensor responses. A closely related metric is thus the dynamic range (DNR), which depends on the FWC normalized by the read noise:

$$DNR = 20 \cdot \log\left(\frac{FWC}{read\ noise}\right) \qquad \text{(eq. 16)}$$

typically expressed in decibels (dB). High DNR value allows working below the saturation even in conditions of long exposure times and/or intense sensor emission.

Note that the brightness of the luminescent sensor and the expected experimental conditions (e.g., sensor concentration, expected exposure times, maximum excitation power density) determine which parameter should be prioritized in choosing the sensor. If pixel saturation is unlikely even for maximum sensor exposure times, the system is *light limited*: read noise and *QE* should be prioritized to maintain good SNR values. For systems where intense signals are to be expected, sensors with high FWC values are preferred to improve SNR, e.g., by increasing the exposure time. The system is said to be *shot noise limited*. Ultimately, this interplay between integration time and noise level enforces a hard SNR floor, driven by noise uncertainty and background variability, that no downstream processing can fully overcome.[139]

Specifically for cameras (rather than, e.g., line array detectors), beyond the above considerations related to the noise level, selecting a specific detection system dictates the achievable spatial resolution, which is fundamentally determined by the interplay between pixel size and the properties of the objective lens (see below). Furthermore, the camera inherently limits temporal resolution by its intrinsic frame-per-second (fps) rate and shutter modality. Global shutters read all pixels simultaneously and are therefore preferred for rapidly evolving processes, avoiding motion artefacts and image distortion (e.g., "jellying"). While rolling shutter sequentially read rows of pixels and are generally sufficient for most static or slowly-changing/-moving specimens.

**Table 4.** Typical performance characteristics of detector technologies used in intracellular luminescence thermometry.

| Detector type | Read noise ($e^-$) | Dark noise ($e^-$(s·pixel)) | QE (%) | Advantages | Typical use |
|---|---|---|---|---|---|
| CMOS | < 2.5 to <7.0 | 0.035-0.25 | 50-85 | Low cost, fast acquisition | Routine widefield imaging |
| sCMOS | 0.7 to 1.6 | 0.001-0.1 | 60-95 | Low noise, high speed, large FOV | Live-cell imaging, multispectral imaging |
| CCD | 2-10 | 0.001-0.05 (cooled) | 40-90 | Excellent linearity, long integrations | Spectroscopy, hyperspectral imaging |
| EMCCD | <1 | 0.001-0.01 | >90 | High sensitivity | Low signal condition |
| PMT | N/A | 10-1000s depending on PMT | 15-45 | Single-photon detection, high temporal resolution | FLIM, TCSPC |

**Point detectors.**[140–144] As mentioned above, APDs and PMTs do not intrinsically provide spatial nor spectral information. Scanning-based microscope architectures or detector arrays are required to generate images. Of more relevance to the field of intracellular luminescence thermometry is the fact that they are used in time-correlated single-photon-counting (TCSPC) techniques, enabling FLIM.

In APDs the dominant noise sources are photon shot and dark noises, alongside the "excess noise" following from the multiplication/avalanche process. IN PMTs too, shot photon and dark noise are the primary sources of noise, while readout noise is practically negligible. Unlike APDs (see below) the multiplication process yields negligible noise.

Similarly to area detectors, the photon shot noise can be modelled according to the following equation:

$$\sigma_S = A\sqrt{F(QE)Nt} \qquad \text{(eq. 17)}$$

Where *A* is the internal detector gain (*G* for PMTs, *M* for APDs) and *F* is the excess noise factor stemming from random secondary emission (in PMT) or impact ionization (in APD). Note that for APDs, the excess noise factor, *F(M)*, is a function of *M* (the avalanche multiplication factor) given the statistical variability in the number of output carriers generated. The dark noise is instead related to thermally produced charge carriers (APDs) and thermoionic emission (PMTs). It is described by:

$$\sigma_D = A\sqrt{F_D Dt} \qquad \text{(eq. 18)}$$

Where $F_D$ is approximately equal to *F*, and *D* are the primary dark electrons/counts.

Like for area detectors, lowering the detector temperature allows substantially reducing the dark noise. This is particularly true in PMTs, where very low values of dark current can be achieved with, i.e., liquid-$N_2$ cooling, hence yielding extremely high values of SNR and sensitivity. On the other hand, APDs are less expensive, sturdier, and more compact, making them more accessible and flexible.

Furthermore, concepts such as full well capacity are not directly applicable. Instead, saturation and non-linear response occur when photon fluxes become sufficiently high to generate space-charge effects within the dynode chain or in the semiconductor material. Important parameters for defining the safe operating range for PMTs and APDs are also the maximum operating current and the dead time ($\tau_D$). The latter is especially relevant in photon-counting applications and is defined as the minimum time separation between two events (photon reaching the sensor) such that both can be detected as separate counts. The maximum counts rates are thus capped by the dead time and roughly equal to its inverse ($1/\tau_D$). In practical terms, for reliable extraction of lifetime values as readout features, low-to-moderate photon rates and gain values (i.e., applied voltage), as well as cooling the detectors.

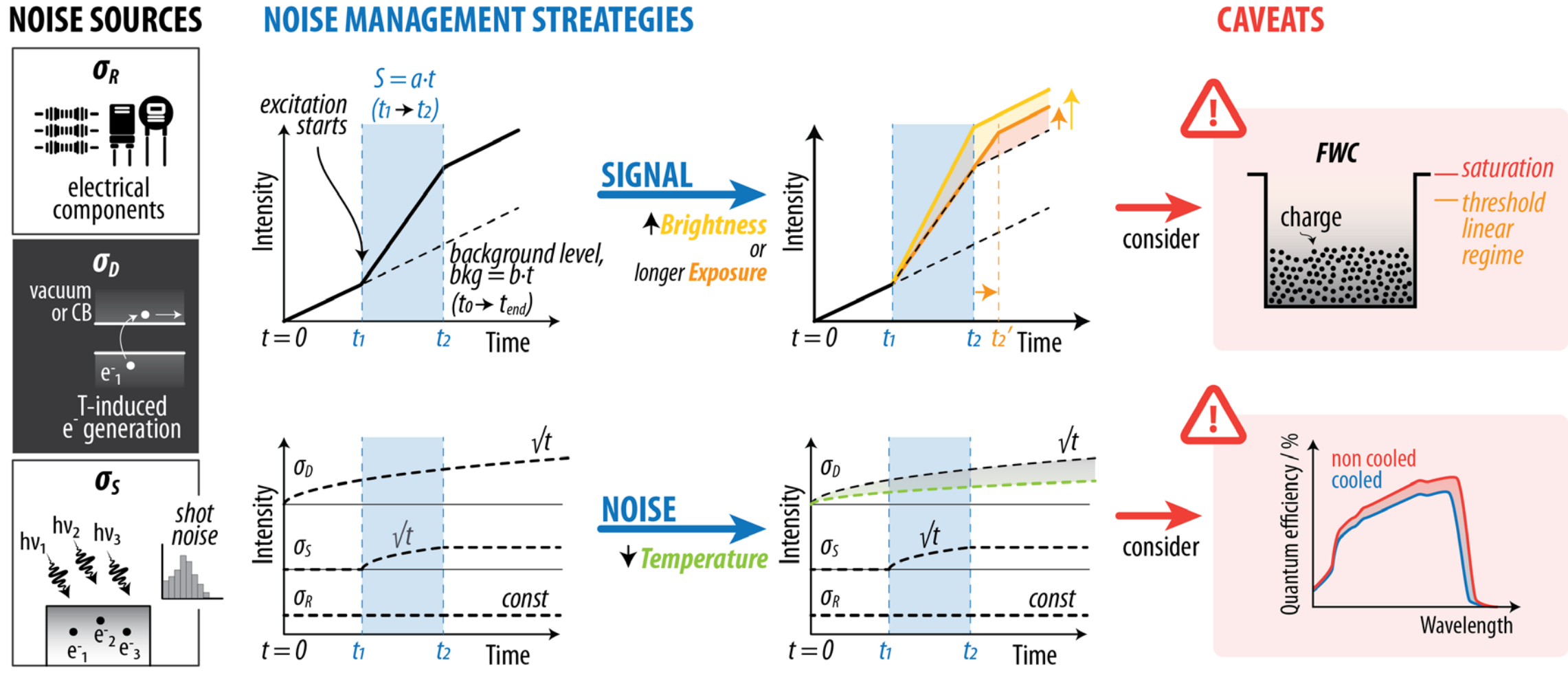

**Figure 8.** Representation of the main noise sources in detectors and how to approach them.

### 4.4.2. Spatial resolution

Depending on the microscopy approach selected, the spatial resolution that can be achieved depends on the optics and type of sensor involved.

In widefield microscopy (and its declination described above), the spatial resolution is only limited by the point spread function (PSF) of the emitting species. The PSF represents how blurry a point-like emitter appears at the detector due to diffraction effects (Airy disk), and its FWHM gives the spatial resolution. Given a wavelength and a numerical aperture of the objective

$$NA = n \sin \alpha \qquad \text{(eq. 19)}$$

(where *n* is the medium's refractive index and alpha is the half-angle of the maximum cone of light accepted by the objective's lens), the lateral resolution ($d_{xy}$) is given by the Rayleigh criterion:

$$d_{xy} = \frac{0.61 \cdot \lambda}{NA} \qquad \text{(eq. 20)}$$

Note that the *NA* value of an objective is fully taken advantage of only when the objective is fully filled. In confocal microscopy and FLIM, the spatial resolution is determined by the PSFs of the excitation and the emission signals, as well as the size of the pinhole—which "cleans" the blurry edges of the emission. As such, the lateral resolution approaches values that are 1.4-fold smaller than in the case of widefield illumination

$$d_{xy} = \frac{0.4 \cdot \lambda}{NA} \qquad \text{(eq. 21)}$$

Besides the improvement in lateral resolution, confocal microscopy also provides substantially enhanced axial resolution owing to the rejection of out-of-focus light by the pinhole. The axial resolution ($d_Z$) can be approximated as:

$$d_Z = \frac{1.4 \cdot \lambda \cdot n}{NA^2} \quad \text{(eq. 22)}$$

where $n$ is the refractive index of the immersion medium and $\lambda$ is the wavelength of the detected emission. Improved axial confinement also reduces contamination from out-of-focus fluorescence originating from other cellular compartments, thereby increasing the reliability of the measured temperature. While eq. 22 provides a good approximation of the axial resolution, a more thorough calculation requires considering also the diameter of the pinhole.[145]

The size of such smallest resolved feature projected on a camera sensor depends on the total magnification (*M*) imposed by the optics (including the objective):

$$d_{sensor} = d_{xy} \cdot M \quad \text{(eq. 23)}$$

Here is where the pixel size/pitch (*p*) becomes relevant, since the Nyquist Sampling Theorem should be satisfied: the smallest optical feature should be sampled with approximately 2.3 pixels. Hence, the sampling limit is given by

$$p \leq \frac{d_{sensor}}{2.3} \; or \; p \leq \frac{(0.61 \cdot \lambda \cdot M)}{2.3 \cdot NA} \quad \text{(eq. 24)}$$

Pixel pitches below this limit result in oversampling, generating larger image files without providing additional spatial information, whereas larger pixel pitches lead to under sampling and pixel-limited resolution. Consequently, the choice of detector pixel size should always be considered together with the magnification and numerical aperture of the microscope objective.

## 5. Calibration: The cornerstone

To use of a luminescent species as an actual luminescent thermometer requires careful calibration. This process entails the acquisition of the luminescence signal (spectra or decay curves) of the probe at different values of temperature in the temperature range in which the probe is expected to be used. Analysis of this calibration dataset provides the values of *F* at each temperature value, so that a calibration curve can be built by fitting the calibration datapoints using a function:

$$F = f(T) \quad \text{(eq. 25)}$$

This curve is later used to obtain the temperature readout during experiment. When a specific value of *F* is observed, inversion of equation 21 provides the thermal readout:

$$T = f^{-1}(F) \quad \text{(eq. 26)}$$

The ideal luminescent thermometer is the one whose performance (in terms of precision and, ideally, accuracy too) is consistent in controlled calibration conditions or during the actual experiment.[3] For that, it is essential to follow standardized and truly quantifiable methodologies, while it is also desirable—whenever possible—to know the physical mechanism underpinning the temperature dependence of the luminescence signal. Knowledge of the physical mechanism offers a powerful tool during the calibration datapoint fitting phase, while also providing a means to highlight possible inconsistencies in the thermometer response that might go unnoticed with a purely mathematical treatment. In addition, artifacts can severely impact the readout. **Section 7** deals with this aspect.

Starting from the considerations made until this point, user-friendly guidelines are provided in this section in an effort to simplify and demystify the calibration process (**Figure 9**).

## 5.1. Standardize experimental conditions before calibration

Calibration of the luminescent thermometer requires attentive selection of the experimental conditions. The first question the researcher should ask themselves is: *what are the conditions in which I will be working during the experimental measurements?* Based on the answer, the calibration conditions should be chosen, since calibration should be performed at conditions that most closely resemble the experimental ones. This approach ensures reliability of the thermal readout.

### 5.1.1. Experimental setup

One of the most important aspects is consistency in the use of the experimental setup. To that end, calibration "in cuvette" cannot be directly translated to a microscope. Differences in the excitation source, detection system, and the whole optical path are bound to introduce changes in the luminescence signal, which, if unaccounted for, are a source of bias. As such, while calibration in cuvette might be a good first screening, it is generally recommended that the final calibration is run at the microscope itself. If calibration at the microscope cannot be carried out, careful characterization of the optical elements should be performed and a "transfer function" should be applied to account for spectral distortions. This is a very important factor that is often overlooked, and that might fuel erroneous conclusions.

For calibrations on a microscope, dedicated systems or modules must be employed to strictly control temperature and isolate other potential sources of cross-sensitivity. For example, stage-top incubators allow to adjust values of temperature and gas concentrations (e.g., $O_2$ and $CO_2$), while also allowing to work with multi-well plates where the luminescent thermometers can be dissolved/dispersed in media with, e.g., varying values of pH, ionic strength, and concentration of relevant biomolecules, all of which are possible sources of cross-sensitivity. Local values of temperature can be picked up by thermocouples inserted in the wells, hence obtaining ground truth values.

For hyperspectral and FLIM modalities, the use of the exact same setup for calibration and measurements is straightforward, since the instruments are designed to acquire spectra and decay curves at each image's pixel. If a thermal readout modality has been chosen—such as luminescence ratio—that requires multispectral widefield acquisitions, the spectral information should be collected before performing imaging. This is generally achieved by replacing the camera with a system for spectral acquisition (as described in the **Section 4**). Before analysis, the calibration dataset collected in this configuration should be corrected to account for the different wavelength-dependent response of the spectral acquisition system and the imaging camera, stemming from different *QE* curves of the detectors and losses in the dispersing and focusing optics.

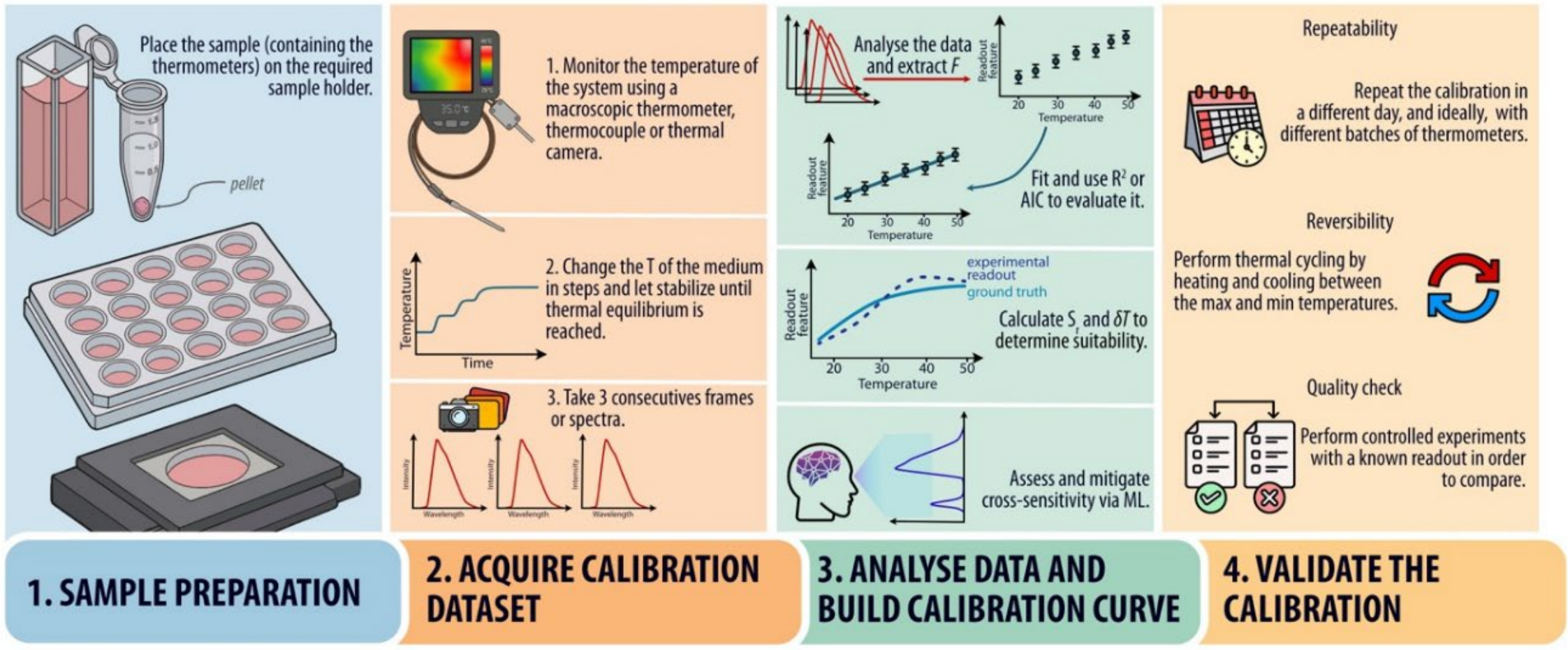

**Figure 9.** Suggested workflow for proper calibration of the luminescence thermometry approach.

### 5.1.2. Temperature range, step size, thermalization time[31]

The temperature range for the calibration should fully cover and slightly exceed the expected operating range. This ensures that no extrapolation is performed during temperature readout in the experiment. For cellular applications, the calibration is typically performed in a range covering physiological temperatures (e.g. 20 to 50 ºC). The step size (temperature difference between calibration points) is often set between 1 and 5 ºC. Given the relatively narrow temperature range of interest in biological systems, it is unlikely that abrupt variations in *F* are encountered. As such, sampling of too-closely spaced temperature conditions is unnecessary, since fitting of the calibration datapoints is carried out. We argue that 5 ºC might be sufficient in most cases.[3,146] That is unless the physics of the luminescent probe (e.g., phase transition materials)[147] suggests that sudden variations are to be expected. Moreover, denser sampling of temperature conditions is also useful to increase the dimensionality of the calibration dataset if ML algorithms are expected to be used in the analysis.

The acquisition of the luminescence signal at each temperature point should be performed in a situation of thermal equilibrium, which is generally guaranteed by thermalization times whose duration depends on the geometry of the system. Thermalization times between 5 and 15 min are usually required, and the external probe used to extract ground truth temperature values is used to verify the stability of the temperature.

### 5.1.3. Excitation intensity

Excitation power density is a critical parameter from several viewpoints. On the one hand, high excitation power densities can lead to photobleaching and photodegradation of the probes, while also contributing to local heating via light-to-heat conversion by molecules in the media and the luminescent thermometers themselves. For this reason, low power densities should be used, which should always be lower than the ones observed to induce photodamage to the cell lines under study. Note that some local heating due to light-to-heat conversion of the luminescent thermometer is generally unavoidable. This is because most luminescent species have PLQY lower than 1. Even when unity PLQY is approached, some spectral mismatch is present between absorption and emission wavelengths (also known as, *quantum defect*), which is a consequence of non-radiative relaxation events that populate the emitting excited state. Both non-unity PLQY and spectral shift contribute to heating. If the illumination conditions are kept constant, this heating introduces a consistent bias in the readout.

On the other hand, changes in excitation power density also yield variations in emission intensity, which are a source of bias in intensity-based thermometry or even in ratiometric approaches that make use of non-linear emitters (i.e., upconverting thermometers) where the ratio is between emission bands with different photon orders.[3]

Therefore, two things should be taken into consideration:

i. The excitation power density (and beam profile, if using, e.g., laser illumination not covering the whole FoV) should ideally be the same for both calibration and subsequent measurements.
ii. One should work within power ranges of a specific excitation wavelength that avoid/minimize heating. This value should be assessed on a case-by-case basis with the help of external sensors and/or simulations.

### 5.1.4. Absorption phenomena

Wavelength-dependent photon absorption by species in the medium and by other luminescent thermometers can lead to bias in temperature readout for spectra-based modalities. During the calibration, this is controlled by keeping the thermometer concentration low and avoiding collecting the luminescence through a thick layer of medium.

If the medium is particularly absorbing—for example, some culture media (like DMEM) contain coloured molecules—optical characterization should be carried out and, for spectra-based approaches, thermometers that do not emit in the absorption regions should be preferred. Alternatively, spectral corrections should be considered during the calibration (and measurement!) steps.[3]

## 5.2. Choose and match the calibration environment

It has been reported that the thermometric response is often affected by surrounding medium and surface-related nonradiative processes.[148] Therefore, the calibration should be performed in an environment that mimics the final measurement conditions as closely as possible.

Because most available luminescent thermometers are susceptible to chemical (e.g., pH, $O_2$, salts) or physical (e.g., mechanical, rheological, electrical) fluctuations in addition to temperature, mapping out the multi-dimensional space of interfering variables is strictly necessary to ensure that experimental results are accepted by the broader community. It is fundamentally meaningless to perform intracellular measurements if these cross-sensitivities and sources of error are not systematically accounted for and corrected[1].

To establish a robust calibration dataset, we suggest a progressive, bottom-up approach rather than immediately calibrating inside the cellular matrix:

1. **Simple media exploration**: Calibration should initiate in a controlled, simple medium (e.g., phosphate-buffered saline, PBS). Here, the isolated effects of primary physiological parameters such as pH, ionic strength, specific ionic species ($Na^+$, $K^+$, $Ca^{2+}$), and oxygenation levels, must be evaluated.
2. **Complex biologically-relevant media**: Once the nanothermometer demonstrates sufficient robustness against these fundamental chemical variations, the system should be tested in more complex media containing biological macromolecules (e.g., proteins, lipids).
3. ***In vitro* validation**: Only after these baseline assessments, the protocol can transition into cellular environments, starting with fixed cells and ultimately moving to live cells.

On the other hand, the field remains deeply divided on whether to rely on calibration datasets obtained from fixed or live cells, as both methodologies present distinct trade-offs:

- **Fixed cells**: This approach offers a stable environment devoid of active metabolic processes. However, it raises critical questions regarding how the chemical fixatives alter the native intracellular chemistry, the structural environment, and even might interact with the thermometers.
- **Live cells**: Calibrating in vivo or in vitro in living systems guarantees an environment that closely mirrors actual measurement conditions. Nevertheless, living cells actively maintain homeostasis and metabolic activity. As such, the temperature set in the culture medium (picked up with a submerged thermocouple) might not be representative of the temperature experienced by the thermometers in the cells. the one.

Hence, calibrating in fixed cells should guarantee a closer set-vs-experienced (by the thermometer) temperature values. This comes at the cost of a physicochemical environment possibly resembling less the one of the final measurement. Calibration in live cells should guarantee the same type of environment experienced by the thermometers during operation. Yet, the normal physiology of cells can introduce variability that is impossible to control. To that end, if the thermometers target organelles/compartments whose temperature is affected by, e.g., metabolic activity, calibration in live cells might suffer from strong deviation between the temperature set and the one experienced by the thermometers. On the other hand, it is unlikely that the cell acts to counter external heating. In fact, homeostatic mechanisms focus primarily on damage control and mitigation of the internal thermal generation rather than active cooling (for which, to the best of our knowledge, there is no direct evidence reported). These mechanisms include *i)* preventing protein denaturation through the action of heat shock proteins[149], *ii)* altering the lipids of the membrane composition,[150] and *iii)* antioxidant production to neutralize reactive oxygen species (ROS) released by the mitochondria under high heat stress.[151] As such, it is unlikely that external changes in temperature are contrasted by active cooling processes taking place within the cell. In addition to the above, accurate calibration of intracellular thermometers in live cells remains challenging because cell movement, probe redistribution, and ongoing metabolic processes during the calibration period can alter the measured optical response. Note that it is also possible to use cell pellets—obtained by aggregation of the cells incubated with the luminescent thermometers—for the calibration step. This approach provides a more stable and reproducible environment by minimizing sample movement and enabling bulk spectroscopic measurements under controlled conditions, thereby improving calibration reliability.[152,153]

While no ultimate strategy is currently laid out for reliable calibration, it is understood that undeniable proof of the reliability of a thermometric approach is the match between the calibration curves obtained in *i)* a cell-free cuvette (to establish the baseline photophysical response of the probe), *ii)* fixed cells (to preserve the complex intracellular matrix without the interference of active metabolic responses), and *iii)* live cells (to capture real-time cellular dynamics).

### 5.2.1. Working with live cells

When working with live cells, a stage-top incubator is generally employed. This is a chamber mounted on the microscope's stage and that allows controlling the composition of the atmosphere (through pre-chamber gas mixing) and temperature experienced by the cells.

Some important aspect to consider when working with live cells are the following:[154,155]

1. Incubation of the cells with the thermometers should be carried out at a concentration previously determined via cytotoxicity tests.
2. An incubation time long enough for effective internalization/targeting (e.g., mitochondria, nucleus, surface) should be allowed, which should be determined via preliminary internalization studies.
3. The cells should be kept on the microscope with environmental control (i.e., gas and medium composition appropriate for the cell line under study).
4. Calibration and measurements in live cells are both affected by cell movements, which should be compensated for during the image analysis.

Note that these considerations are also relevant when carrying out the thermometric readouts during the experiments.[25,3]

## 5.3. Construct and validate the calibration curve

1. **Analyse the calibration dataset**: One the dataset has been acquired; it should be analysed to extract the best readout feature (*F*). This task can be approached in several ways:

   a. if the physical model underpinning the thermal sensitivity is known, the, e.g., spectra or decay curves should be fitted accordingly hence extracting the readout feature
   b. a phenomenological trial-and-error approach can be undertaken. For example, the signal over two arbitrary spectral ranges that appear to have opposite trends as a function of temperature can be integrated and used to obtain a LIR.
   c. ML algorithms can be used to extract readout features more complex in nature, such as involving highly non-linear spectral combinations. ML algorithms are recommended to process complex, multi-dimensional inputs, particularly for recognizing and untangling overlapping signal features that analytical models struggle to separate.[156,157]

Note that this step is critical, and it requires proper and consistent data pre-processing (e.g., normalization, denoising…).

2. **Plot and report datapoints**: Once $F$ has been defined, the scatter calibration curve should be plotted ($F_i$, for each $i$-th temperature value). Crucially, these datapoints must also be reported as a function of the different values of tested interfering parameters and varied calibration environments. This allows for a thorough investigation of potential cross-sensitivity effects.

3. **Fit** to the chosen model and **extract** parameters: Fit the data to the chosen physical/analytical model.

4. **Evaluate performance** (see **Section 6** for further details about this):

- Goodness of fit: Use metrics such as R-squared or the Akaike Information Criterion (AIC).[158,159]
- Uncertainty and sensitivity: For conventional models, calculate the relative thermal sensitivity $S_r$ and the predicted temperature uncertainty $\delta T$. Compare $\delta T$ against application requirements (e.g., 0.1 °C for many biological processes) to determine suitability.[139,160]
- Assess (and mitigate) cross-sensitivity via ML: Use metrics such as CST and, ideally, associated error plots to quantify the effect of cross-sensitivity. Where cross-sensitivity to physiochemical parameters is present, ML algorithms can be leveraged to overcome these confounding factors. By training models on multi-parameter datasets, the ML algorithm can isolate the temperature-dependent response from background interference, effectively bypassing traditional cross-sensitivity limitations.

5. **Validate the calibration**:

- Repeatability: Repeat the entire calibration protocol on a different day. Ideally, also different batches of thermometers (either purchased or synthesized) can be tested to guarantee probe reliability.
- Reversibility experiments: Perform thermal cycling by heating and cooling between the minimum and maximum T values to evaluate hysteresis.
- Quality check: Perform controlled experiments with a known, predictable temperature readout (see below).

The last aspect is pivotal to transition a luminescent thermometer from a controlled calibration phase to a reliable real-time sensing application. A standardized quality check is not currently available, yet we argue that a dynamic validation should be preferred, where temperature changes in a predictable and well-characterized manner both in terms of values and characteristic timespan. As a first approach, for calibrations carried out in cuvettes and in fixed cells, a quality check can be performed varying the temperature with a heating element and obtaining a continuous readout via luminescence thermometry.[31] The temperature values so obtained should be compared with the temperature values obtained by a thermocouple placed in the proximity of the emission collection point and, ideally, with the results obtained from simulations of heat diffusion.[161] Yet, real validation of the calibration should be ultimately performed in conditions close to the ones expected during the experiment to be carried out. We provide suggestions regarding this aspect in **Section 7**.

### 5.3.1. Machine learning-assisted calibration

A comprehensive discussion about how to use ML in luminescence thermometry goes beyond the scope of this manuscript. A dedicated manuscript is currently in the works. Note that the library of ML algorithms is vast, and general guidelines cannot be easily provided. Yet, for the purpose of the discussion at hand, the following general considerations can be made.

When ML approaches are employed for thermometric calibration, particular attention must be paid to the design and validation of the calibration dataset. To guarantee effective algorithm training, the dimensionality of the calibration dataset should substantially exceed the number of effective variables retained by the model.[162,163] The minimum dimensionality of the calibration dataset—which is intimately correlated with the total number of spectra or decay curves recorded during calibration—strongly depends on the type of algorithm to be trained; yet the higher the number of spectra/decay curves, the better. For example, a linear DR algorithm such as PCA might be satisfied with a few hundred spectra, acquired as replicates over the sampled temperature values.[31] Complex algorithms such as convolutional neural networks (CNNs) require instead thousands of spectra/decay curves to meaningfully extract non-linear relationships.[164]

Calibration data should be separated into independent training and validation datasets. Approximately 80% of the data is often used for training and 20% for validation. Shuffling of the datasets is useful to avoid overfitting and is encouraged both before every training epoch within the training dataset, and after a predetermined number of epochs between the training and validation datasets.[165] Progress of the training is monitored via analysis of the loss function (generally a variation of the average absolute or squared error, MAE or MSE), whose value should decrease during the training.[166–169] Independent testing (quality check) is more reliably carried out using photoluminescence data acquired at values of temperature (and of possible interfering parameters!) that are different from the used selected for training and validation. Whenever possible, model performance should be evaluated using measurements acquired from independent experiments, different samples, or different days rather than random subsets of the same dataset used for training and validation.[170,171]

Finally, the complexity of the selected ML model should remain commensurate with the size and quality of the available dataset. In many thermometric applications, simpler and more interpretable models often provide more robust and reproducible temperature predictions than highly complex architectures that may be prone to overfitting. Whenever possible, ML-based predictions should be benchmarked against conventional calibration approaches and accompanied by estimates of prediction uncertainty.

## 5.4. Reporting calibration for reproducibility

To ensure reproducibility and transparency, the underlying source code for all algorithms must be deposited in a public repository and made freely available. For that, after performing the calibration, explicitly report:

- Excitation wavelength, power density, and beam geometry.
- Optical setup (detector type, objective, lenses…), settings, and spectral window.
- Sample environment (medium composition, pH, presence of live and dead cells).

- Temperature control method, reference thermometer, and thermal equilibration times.

This level of detail is essential for reproducible, quantitative luminescence thermometry and for the field's standardization.[1,3]

# 6. Data analysis and handling

## 6.1. Data analysis

Following data extraction, fitting of the calibration experimental points (*F*-vs-*T*) should be carried out. Evaluating the goodness of fit for these calibration curves requires robust statistical indicators:

- **The Coefficient of Determination ($R^2$)**: Measures the proportion of variance in the dependent variable that is predictable from the independent variable—hence, it indicated the goodness of fit. A value closer to 1 indicates that the regression model fits the data points well.[172]
- **The Akaike Information Criterion (AIC)**: Evaluates the relative quality of statistical models for a given set of data. Unlike $R^2$, AIC does not consider the variance described by the model and it penalizes model complexity (the number of estimated parameters), helping researchers avoid overfitting and select the most parsimonious fitting function. AIC is mainly useful when the fitting function is not known a priori, i.e., the physical mechanism relating *F* to *T* is not known.[173,174]

## 6.2. FAIR practices[22,75,80,175–178]

Intracellular thermometry generates complex, heterogeneous datasets that will need to rely heavily on FAIR (Findable, Accessible, Interoperable, Reusable) data supervision. This work showed that bias from cell activity and probe interference can compromise reliability.

Therefore, reporting raw data, calibration procedures, metadata, and analysis code is therefore essential to allow independent re-analysis of contentious temperature map, compare probes, calibration strategies, and analysis pipelines across laboratories, and support meta-analyses of probe bias and performance in different biological contexts.

In **Table 4**, FAIR challenges and strategies proposed to overcome them are displayed:

# 7. Conclusions and emerging directions

In this manuscript, we have provided a holistic view of the field of intracellular luminescence thermometry, implementing physical, chemical, biological, and mathematical considerations. We have especially attempted to capture critical aspects related to interpretation of the results, thermometric approach design, and consistent data reporting. We have favored a didactical approach, without shying away—when deemed convenient—from deeper dives into some more technical aspects. In so doing, we have highlighted how luminescence thermometry has made significant strides towards monitoring real-time thermal dynamics at the cellular and even sub-cellular level, yet several criticalities persist.

**Table 4.** Challenges and strategies for FAIR practices.

| PROBLEM IN INTRACELLULAR THERMOMETRY | WHAT THIS MEANS? | WHAT YOU CAN DO (FAIR STYLE) |
|---|---|---|
| Temperature readings are biased by cell activity and probe behaviour. | The probe signal changes not only with temperature, so maps may be wrong. | Share full calibration data, and include uncertainty so others can judge reliability. |
| Many instruments, many formats, no common structure. | Each lab's produce different files and metadata, being hard to combine or reuse. | Use a common data structure (e.g. JSON based on metrology frameworks) and "digital twins". |
| Important details about probes and experiments are missing. | Published work often lacks enough information to reproduce results. | Record and share rich metadata following FAIR guidance. |
| Data are hard to find, access, and reuse later. | Other groups cannot discover or analyse them. | Deposit raw and processed data with persistent IDs and standard vocabularies. |
| Ethical and community norms are unclear. | Different fields have different views on sharing and reusing data. | Combine FAIR with community rules that make sharing worthwhile. |

The tension between theoretical and experimental results is worrying, and the defendants of each position are resolute in standing behind their arguments. The key to bridging the two perspectives lies in proposing a truly reliable intracellular luminescence thermometry approach. Easier said than done, achieving this goal requires a concerted effort in terms of probe design, fine understanding of cell physiology, and full control and knowledge of the optical setup employed for conducting the measurements.

Despite several reports claiming reliable intracellular thermal readouts, doubts linger regarding possible sources of cross-sensitivity. To appropriately tackle them, we consider necessary to sweep the multidimensional space described by varying stimuli in cells or specific organelles. Depending on the nature of the thermometer, it will accumulate in different cell compartments, which in turn will determine the range of variability and the nature of the interfering parameters (e.g., pH, viscosity, salt and sugar concentration). As such, this scan should be cell- or even organelle-specific. This is a tedious-yet-essential task, which will help in assessing precision and accuracy in thermal readouts. This thorough scan is also at the basis of what we consider a powerful emerging tool in luminescence thermometry: machine learning algorithms for spectral analysis. Properly trained, these algorithms are unmatched in terms of simultaneously improving readout precision and accuracy. Like all powerful tools, these algorithms should be handled with care, since mathematical artefacts such as overfitting are a real possibility, and they are bound to compromise the reliability of the readouts.

We cannot offer a final verdict regarding how calibration should be performed, but we suggest that at the very least fixed cells should be used. The use of live cells for calibration seems attractive mainly for thermometers residing in the cytosol, accumulating at the external membrane, or residing within endosomes. Match between the calibration curves obtained in fixed and live cells is ultimate proof of the reliability of the thermometer, but we argue that a mismatch in those results should not be considered as clear evidence of lack of reliability of the thermometric approach.

Rather, standardized validation experiments should be identified and run after calibration. Following a literature survey, we suggest that an approach can be the measurement of the temperature increase induced by specific external stimuli applied in a deliberate manner (**Figure 10**). These are both chemical and physical in nature, such as the addition of glucose and oleic acid to the cell culture medium, or illumination with a laser. It goes without saying that all these tests should be first standardized in terms of cell type, concentration of the added chemical species, and power density, beam size, wavelength, setup geometry, and duration of the illumination. This is ultimately a metrology problem, and we believe the community is taking the right steps towards this direction.

To that end, an effort to standardize the field requires a concerted, global effort between leading groups. Identification of promising thermometers and definition of standard operating procedures (SOPs) to run quality checks and the final experiments is mandatory for gaining the trust of the several communities involved in the study of cellular physiology.

Once the above aspects are addressed, reliable thermal readouts can be performed. In this vein, some of the avenues we consider exciting for further exploration are the following.

**Solution of the longstanding controversy about mitochondria's temperature.** As banal as it might sound, undoubtedly the first experiment one would want to perform with a reliable thermometer designed to target mitochondria is monitoring thermogenesis in those organelles. Cell lines of interest include different cancer cell lines (e.g., HeLa, MCF-7, HepG2, PC-3, U-87 MG) and adipocytes (e.g., hTERT-immortalized and hMADS), where thermogenesis is expected to be substantial, due to increased metabolic activity. Studies on cancer therapy as well on eating disorders would greatly benefit from accurate mapping of temperature at the cellular level.

**Identification of thermal patterns in the life cycle of cells.** It is currently unknown whether specific temperature patterns are recognizable during specific checkpoints of the cell life. Meiosis, mitosis, and cytokinesis are some of the phases of more interest. It is well documented that changes in external temperature do affect these processes, but localized temperature changes have not been investigated to the best of our knowledge.

**Investigation of non-adherent cells and 3D spheroids.** Current intracellular temperature measurements are mostly relegated to standard, two-dimensional adherent cell cultures. To elucidate how the supporting substrate and microenvironment influence thermal dynamics, the study of non-adherent cells and 3D cellular spheroids is key. This approach will clarify the potential thermal dissipation or buffering effects caused by the substrate, as well as the physiological relevance of continuous physical contact with the substrate.

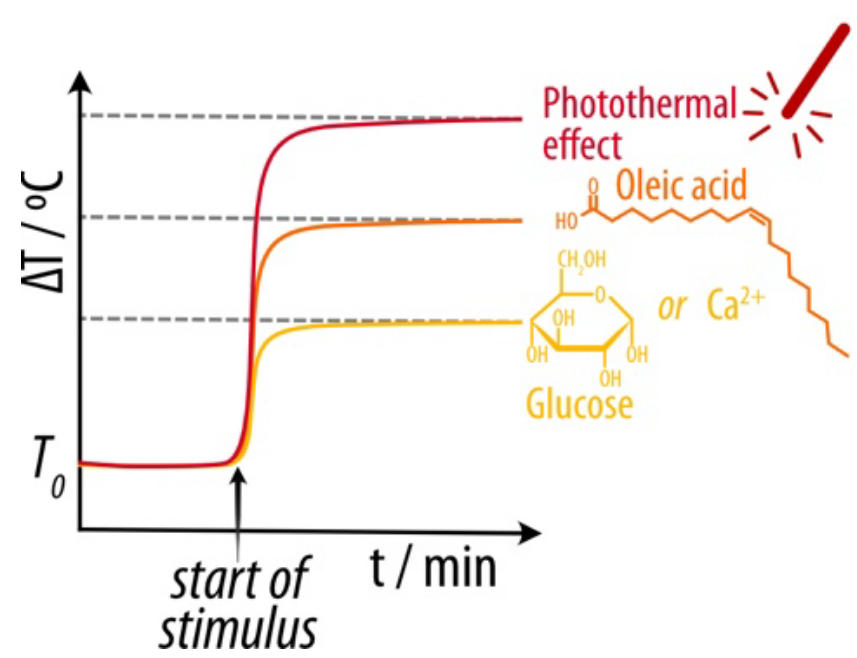

**Figure 10.** Representation of temperature variations recorded when adding different stimulus to the cells.

**Implementation of automation in intracellular luminescence thermometry.** Several fields are introducing the concept of autonomous laboratories, where the operator factor is minimized. Luminescence thermometry is likely to benefit too from this automatization at several steps of the process. On the one hand, luminescence thermometer preparation using robots ensures greater material reproducibility and, possibly, better sterility. Incubation with cells and AI-driven observations at the microscope are also conceivable.

## Acknowledgments

Funded by the European Union (ERC, MAtCHLESS, 10116287). Views and opinions expressed are however those of the author(s) only and do not necessarily reflect those of the European Union or the European Research Council. Neither the European Union nor the granting authority can be held responsible for them. This work was financed by Grant PID2023-146775OB-I00 (INCLINA) funded by MCIN/AEI/ 10.13039/501100011033, by the Instituto de Salud Carlos III (PI24/01535), by the Comunidad de Madrid (S2022/BMD-7403 RENIM-CM) and co-financed by the European structural and investment fund. This work has been also supported by LUMETH. The project (25HLT06 Lumeth) has received funding from the European Partnership on Metrology, co-financed from the European Union’s Horizon Europe Research and Innovation Programme and by the Participating States.